\documentclass[fleqn,usenatbib]{mnras}

\usepackage{newtxtext,newtxmath}
\usepackage[T1]{fontenc}
\usepackage{subfig} 

\DeclareRobustCommand{\VAN}[3]{#2}
\let\VANthebibliography\thebibliography
\def\thebibliography{\DeclareRobustCommand{\VAN}[3]{##3}\VANthebibliography}

\usepackage{graphicx}	% Including figure files
\usepackage{amsmath}	% Advanced maths commands

\usepackage{amssymb}	% Extra maths symbols
\usepackage{longtable}

\title[New TESS Heartbeat Star Systems]{Twenty-three New Heartbeat Star Systems Discovered Based on TESS Data}

\author[Li et al.]{
Min-Yu Li,$^{1}$
Sheng-Bang Qian,$^{2,3}$\thanks{E-mail: qiansb@ynu.edu.cn}
Ai-Ying Zhou,$^{4}$
Li-Ying Zhu,$^{1,5}$
Wen-Ping Liao,$^{1,5}$
\newauthor
Er-Gang Zhao,$^{1}$
Xiang-Dong Shi,$^{1}$
Fu-Xing Li,$^{1}$
Qi-Bin Sun,$^{2,3}$
\\
$^{1}$Yunnan Observatories, Chinese Academy of Sciences, Kunming 650216, People's Republic of China\\
$^{2}$Department of Astronomy, School of Physics and Astronomy, Yunnan University, Kunming 650091, People's Republic of China\\
$^{3}$Key Laboratory of Astroparticle Physics of Yunnan Province, Yunnan University, Kunming 650091, People's Republic of China\\
$^{4}$National Astronomical Observatories, Chinese Academy of Sciences, A20 Datun Road, Chaoyang District, Beijing 100101, People's Republic of China\\
$^{5}$University of Chinese Academy of Sciences, No.1 Yanqihu East Road, Huairou District, Beijing 101408, People's Republic of China
}

\date{Accepted XXX. Received YYY; in original form ZZZ}

\pubyear{2024}

\begin{document}
\label{firstpage}
\pagerange{\pageref{firstpage}--\pageref{lastpage}}
\maketitle

\begin{abstract}
Heartbeat stars (HBSs) are ideal astrophysical laboratories to study the formation and evolution of binary stars in eccentric orbits and the internal structural changes of their components under strong tidal action. We discover 23 new HBSs based on TESS photometric data. The orbital parameters, including orbital period, eccentricity, orbital inclination, argument of periastron, and epoch of periastron passage of these HBSs are derived by using a corrected version of Kumar et al. model based on the Markov Chain Monte Carlo (MCMC) method. The preliminary results show that these HBSs have orbital periods in the range from 2.7 to 20 days and eccentricities in the range from 0.08 to 0.70. The eccentricity-period relation of these objects shows a positive correlation between eccentricity and period, and also shows the existence of orbital circularization. The Hertzsprung-Russell diagram shows that the HBSs are not all located in a particular area. The distribution of the derived parameters suggests a selection bias within the TESS survey towards HBSs with shorter periods. These objects are a very useful source to study the structure and evolution of eccentricity orbit binaries and to extend the TESS HBS catalog.

\end{abstract}


%%%%%%%%%%%%%%%%%%%%%%%%%%%%%%%%%%%%%%%%%%%%%%%%%%

%%%%%%%%%%%%%%%%% BODY OF PAPER %%%%%%%%%%%%%%%%%%

\section{Introduction}

Heartbeat stars (HBSs) are a subclass of eccentric orbit binaries, named for their echocardiogram-like light curves \citep{2012ApJ...753...86T}. The amount of deformation depends on the eccentric orbital phase, with the strongest effect occurring during the periastron passage. This tidal variation contributes to the ``heartbeat'' feature near periastron and therefore affects the shape of the light curve. HBSs are ideal astrophysical laboratories to study the formation and evolution of binary stars in eccentric orbits; it is also an important class of objects to study the properties of binaries with eccentric orbits, and to search for special celestial bodies, etc. Therefore, they are important and valuable in scientific research. Although theoretical work on HBSs is extensive, the small amplitude and short duration of the light variations make them difficult to detect with ground-based telescopes, and only a few have been discovered \citep{2017MNRAS.472.1538F}. It was not until the release of long-period, high-precision photometric data from the Kepler space telescope \citep{2010Sci...327..977B} that the tiny changes in the light curves were discovered in large numbers and began to be studied extensively.

After the first Kepler HBS KOI-54 (KIC 8112039) was reported \citep{2011ApJS..197....4W}, other HBSs have been studied based on the Kepler photometric data. Building on the foundational work of \citet{2016AJ....151...68K}, which presented a catalog of 173 Kepler HBSs, a number of investigations have used these objects for further study \citep{2014AA...564A..36B, 2013MNRAS.434..925H, 2016MNRAS.463.1199H, 2018MNRAS.473.5165H, 2016ApJ...829...34S, 2017MNRAS.469.2089D, 2017ApJ...834...59G, 2019ApJ...885...46G, 2020ApJ...888...95G, 2020ApJ...903..122C, 2021MNRAS.508.3967O, 2021ApJ...922...37O, 2022MNRAS.517..437G, 2023ApJS..266...28L}. In addition, \citet{2022ApJS..259...16W,2022ApJ...928..135W} have reported 991 OGLE HBSs, the largest catalog of HBSs to date.

The Transiting Exoplanet Survey Satellite (TESS; \cite{2015JATIS...1a4003R}) was launched by NASA as an exoplanet survey mission in 2018. It has a field of view of 24 $\times$ 96 degrees and observes a portion of the sky for 27 days before turning to a new position. MACHO 80.7443.1718 (TIC 373840312) is one of the most interesting HBS that has been studied in a number of papers \citep{2019MNRAS.489.4705J, 2021MNRAS.506.4083J, 2023A&A...671A..22K, 2023NatAs.tmp..173M, 2024A&A...686A.199K}. \citet{2021A&A...647A..12K} reported a most massive HBS from TESS and 19 other HBSs. \citet{2023AJ....166...42W} studied HBS FX UMa (TIC 219707463). \citet{2022arXiv221210776B} have reported a catalog of TESS variable stars, including 96 HBS candidates.

Thanks to the continued observations by the TESS survey telescope, in this work we have discovered 23 new HBSs based on TESS photometric data. Section \ref{sec:data_modeling} presents a preliminary analysis of the objects, focusing on the determination of their orbital parameters of these objects. Section \ref{sec:properties} delves into the properties of the identified HBSs. This section explores the relationship between eccentricity and orbital period, presents the objects within a Hertzsprung-Russell diagram, and analyzes the distribution of their various parameters. Section \ref{sec:conclusions} summarizes the key findings of this study and draws conclusions based on the analyses conducted in the preceding sections.

\section{The TESS Heartbeat Star Systems} \label{sec:data_modeling}
\subsection{Selection of Heartbeat Stars}
The HBS star systems were discovered by the following steps. First, we searched for the HBS candidates by focusing only on the shape of the light curves. The photometric data were downloaded from the public data processed by TESS-SPOC (the TESS Science Processing Operations Center) and the MIT QLP (the MIT Quick-Look Pipeline), using the {\tt\string lightkurve} package \citep{2018ascl.soft12013L}. If there is a ``heartbeat'' signal in the light curve, the object can be classified as an HBS candidate. After a visual search of about 160,000 TESS objects, about 60 HBS candidates were found.

Second, we visually inspected the light curves and selected data sources with low scatter and significant heartbeat signals for analysis. We also detrended the data using the Locally Weighted Scatter-plot Smoothing (LOWESS) approach \citep{cleveland1979robust}, and then removed obvious outliers. In addition, the TESS sectors contain varying amounts of blending in the data of some of the systems, which may be contaminated by other sources. We also excluded these high blending sectors from our analysis.

Third, we performed a rough fit of these HBS candidates using the approach described in Section \ref{sec:K95plus}. We then excluded some objects with one of the following characteristics: (a) A reliable orbital period cannot be determined because there is no periodicity in the appearance of the heartbeat signal. (b) The heartbeat feature becomes no longer significant in phase-folded light curves. (c) The light curves do not fit properly.

Finally, we obtained 23 new HBSs whose basic parameters are shown in Table \ref{tab:basicparms}. The first column is the TESS ID; columns 2 $-$ 4 are obtained from the SIMBAD website \footnote{\url{https://simbad.u-strasbg.fr/simbad/}}; columns 5 $-$ 7 are derived in Section \ref{subsec:HR}; columns 8 and 9 represent the selected data source and sectors, respectively.

\setlength{\tabcolsep}{1.2mm}{
\begin{table*}
	\caption{Basic parameters of the 23 TESS HBSs.}
	\label{tab:basicparms}
	\begin{tabular}{rcccccccc}
		\hline
		TESS ID & Simbad main ID & Coords (J2000) & V(mag) & $\pi$(mas)  & $T_{\rm eff}$(K) & log(L/L$_{\odot}$) & D. S. & sector(s) \\
		\hline
		21222483 & HD 72965 & 08 36 10.95 +13 46 39.17 & 7.39 & 4.55 & 9773 & 1.78 & (2) & 44$-$46, 72\\ 
		58177975 & HD 283527 & 04 17 28.73 +26 52 13.30 & 9.89 & 2.12 & 7951 & 1.74 & (1) & 43, 44\\ 
		71410794 & BD-22 553 & 03 08 41.21 $-$21 26 11.42 & 10.68 & 2.02 & 6805 & 1.16  & (1) & 4, 31\\ 
		85177749 & HD 282756 & 04 59 46.09 +29 35 19.29 & 9.83 & 3.63 & 9707 & 1.41  & (1) & 19, 43, 44\\ 
		118024242 & HD 72730 & 08 33 54.34 $-$18 26 21.06 & 9.73 & 1.28 & 8047 & 1.82 & (1) & 8, 34, 61\\ 
		137810570 & HD 238047 & 11 50 50.50 +57 39 17.47 & 9.46 & 3.10 & 6661 & 1.17  & (1) & 14, 15, 21, 22, 41, 48\\ 
		153123772 & CD-24 3564 & 05 57 08.03 $-$24 51 43.61 & 10.92 & 0.98 & 7543 & 1.55  & (1) & 6, 32, 33\\ 
		153695072 & HD 43099 & 	06 14 47.85 +07 54 18.58 & 9.0 & 1.46 & 7472 & 1.79 & (2) & 6, 33\\ 
		202007047 & HD 48689 & 06 44 54.10 +07 27 49.11 & 8.57 & 2.60 & 7047 & 1.66  & (1) & 6, 33\\ 
		262036561 & HD 53563 & 07 05 50.76 +08 07 58.85 & 9.62 & 1.98 & 7826 & 1.48 & (1) & 33 \\ 
		265473090 & HD 17173 & 02 47 46.95 +58 38 39.33 & 9.40 & 2.68 & 7095 & 1.28 & (1) & 58\\ 
		283917392 & HD 289757 & 07 13 32.84 +01 48 31.29 & 10.59 & 1.53 & 7082 & 1.44 & (1) & 7, 33\\ 
		292160826 & CD-25 10906 & 15 23 38.02 $-$25 48 33.70 & 10.21 & 2.22 & 6574 & 1.37 & (1) & 11, 38\\ 
		353235026 & BD+10 81 & 00 43 24.47 +11 35 45.20 & 9.29 & 3.24 & 6600 & 1.27  & (1) & 42, 43\\ 
		363674490 & HD 195713 & 20 32 05.63 +25 21 07.41 & 7.91 & 2.65 & 9097 & 1.92 & (1) & 41, 55\\ 
		370209445 & HD 236746 & 01 28 02.67 +57 17 49.09 & 9.06 & 2.73 & 8420 & 1.63 & (1) & 18, 58\\ 
		370269453 & HD 8824 & 01 28 14.72 +56 28 20.82 & 8.81 & 1.94 & 10700 & 2.11 & (1) & 18, 58\\ 
		386138719 & HD 58816 & 07 26 43.85 $-$12 27 30.26 & 8.55 & 3.13 & 8549 & 1.50 & (1) & 7, 34\\ 
		388992242 & HD 121666 & 13 56 49.72 +00 10 58.80 & 9.40 & 3.78 & 6441 & 1.11 & (1) & 50\\ 
		411636838 & HD 8634 & 01 25 35.68 +23 30 41.46 & 6.187 & 13.85 & 6447 & 1.22 & (1) & 57\\ 
		412342944 & HD 57688 & 07 21 35.85 $-$18 09 34.21 & 9.62 & 0.66 & 8287 & 2.43 & (2) & 34\\ 
		444866141 & HD 237497 & 06 30 54.52 +59 34 19.37 & 9.51 & 3.50 & 6514 & 1.12 & (1) & 20, 60\\ 
		452822117 & HD 65778 & 08 00 22.53 +03 14 56.21 & 7.92 & 4.21 & 7413 & 1.50 & (1) & 34, 61\\
		\hline
		\multicolumn{9}{l}{The first column is the TESS ID, columns 2$-$4 are the basic parameters of the objects, columns 5$-$7 are from Section \ref{subsec:HR},}\\
		\multicolumn{9}{l}{column 8 represents the data source as follows: (1) TESS-SPOC data; (2) MIT QLP data,}\\
		\multicolumn{9}{l}{column 9 shows the sector(s) used for analysis.}\\
	\end{tabular}
\end{table*}

\subsection{Modeling of the Heartbeat Star Systems} \label{sec:K95plus}
We fit the corrected version of the \citet{1995ApJ...449..294K} model (K95$^+$ model: Equation (\ref{equation:one})), which has been corrected by \citet{2022ApJ...928..135W}, to the light curve.
\begin{equation}\label{equation:one}
	\frac{\delta F}{F}(t) = S\cdot\frac{1-3\sin^2i\sin^2(\varphi(t)+\omega)}{(R(t)/a)^3}+C,
\end{equation}
The K95$^+$ model contains seven parameters: orbital period ($P$), eccentricity ($e$), orbital inclination ($i$), argument of periastron ($\omega$), the epoch of periastron message ($T_{0p}$), the amplitude scaling factor ($S$), and the fractional flux offset ($C$). The fitting approach is based on the Markov Chain Monte Carlo (MCMC) method with the {\tt\string emcee} v3.1.2 Python package \citep{2013PASP..125..306F}, following our previous work \citep{2023ApJS..266...28L}. Figure \ref{fig:corner} shows the corner plot of the MCMC fitting procedure for TIC 21222483 as an example. For all systems, their average autocorrelation times are in the range of 80 $-$ 90 consecutive steps in the MCMC run.

Finally, the parameters and their uncertainties are derived. Table \ref{tab:hbparms} shows the parameters of these 23 new HBSs. The first column is the TESS ID; column 2 is the orbital period in days; column 3 is the eccentricity; column 4 is the inclination in degrees; column 5 is the argument of periastron in degrees; column 6 is the epoch of the periastron passage in TJD=BJD$-$2,457,000; column 7 is the amplitude scaling factor; column 8 is the fractional flux offset; column 9 is the data source. Figures \ref{fig:hbs1} $-$ \ref{fig:hbs3} show the fit results of the HBSs. In addition, the panel (h) of Figure \ref{fig:hbs3} shows a typical example of a higher amount of blending in sector 7 of TIC 452822117. Therefore, we exclude it from our analysis.

\setlength{\tabcolsep}{1.2mm}{
\begin{table*}
	\caption{Parameters of the K95$^+$ model fitted to the light curves of the 23 TESS HBSs.}
	\label{tab:hbparms}
	\begin{tabular}{rccccccc}
		\hline
		TESS ID & $P$(d) & $e$ & $i$($^\circ$) & $\omega$($^\circ$)  & $T_{0p}$(TJD) & $S$($\times$10$^{-4}$) & $C$($\times$10$^{-4}$) \\
		\hline
		21222483 & 2.653952(21) & 0.31744(46) & 38.153(31) & 136.03(13) & 2503.18032(58) & 46.514(98) & -16.882(85)  \\ 
		58177975 & 6.24509(24) & 0.25302(50) & 54.42(11) & 7.88(16) & 2482.8182(19) & 26.129(71) & -1.269(92)  \\ 
		71410794 & 15.1847(34) & 0.6977(25) & 55.85(65) & 122.9(13) & 1438.69(16) & 0.915(29) & 0.037(73)  \\ 
		85177749 & 3.897662(20) & 0.3240(31) & 32.84(17) & 32.46(78) & 1818.7181(47) & 7.041(96) & -3.662(71)  \\ 
		118024242 & 4.1497773(75) & 0.0822(11) & 23.01(12) & 157.30(29) & 1522.8297(32) & 63.63(73) & -46.19(69) \\ 
		137810570 & 3.8415356(42) & 0.30298(70) & 39.634(57) & 6.22(22) & 1690.9513(13) & 14.915(44) & -4.296(41) \\ 
		153123772 & 11.393982(41) & 0.5853(11) & 60.36(30) & 170.11(46) & 1480.0822(40) & 4.716(39) & 1.12(10) \\ 
		153695072 & 6.348147(16) & 0.18483(60) & 43.947(83) & 135.62(21) & 1478.8890(33) & 56.54(20) & -13.75(18) \\ 
		202007047 & 20.03607(24) & 0.5253(18) & 47.95(23) & 135.38(73) & 1482.983(13) & 1.729(21) & -0.284(36) \\ 
		262036561 & 5.1547(21) & 0.2166(24) & 37.77(23) & 75.43(82) & 2212.0656(94) & 15.28(22) & -6.62(20) \\ 
		265473090 & 4.35710(39) & 0.3414(13) & 54.91(19) & 154.27(40) & 2889.7426(27) & 11.617(64) & 0.182(77) \\ 
		283917392 & 4.987200(35) & 0.3591(34) & 64.6(11) & 49.66(100) & 1493.0868(75) & 3.707(95) & 0.51(10)  \\ 
		292160826 & 4.909607(32) & 0.2466(34) & 29.19(17) & 172.75(85) & 1609.2531(66) & 16.37(26) & -9.35(19) \\ 
		353235026 & 4.00558(18) & 0.3078(10) & 33.997(67) & 121.63(23) & 2452.5997(19) & 19.013(90) & -8.466(78) \\ 
		363674490 & 3.0298284(42) & 0.29944(57) & 45.812(63) & 168.00(18) & 2424.50405(94) & 13.429(34) & -3.552(36) \\ 
		370209445 & 3.7715154(24) & 0.20115(45) & 60.28(18) & 162.46(16) & 1796.0841(14) & 22.848(83) & 2.206(92) \\ 
		370269453 & 5.960762(11) & 0.3194(12) & 40.834(87) & 150.75(30) & 1796.4456(32) & 13.899(70) & -3.670(56) \\ 
		386138719 & 7.186588(85) & 0.2716(58) & 26.53(22) & 29.0(10) & 1494.140(10) & 6.14(15) & -3.96(11) \\ 
		388992242 & 6.6863(65) & 0.3179(32) & 46.54(35) & 52.02(87) & 2676.6461(96) & 7.81(15) & -1.81(14) \\ 
		411636838 & 5.45310(34) & 0.29354(83) & 29.827(37) & 167.58(17) & 2863.5449(13) & 10.749(36) & -5.459(24) \\ 
		412342944 & 15.1357(45) & 0.28077(86) & 69.01(50) & 128.40(31) & 2249.0675(83) & 18.43(15) & 6.81(18) \\ 
		444866141 & 3.1665730(29) & 0.17079(66) & 61.33(30) & 50.91(29) & 1842.8294(21) & 17.51(12) & 2.67(11) \\ 
		452822117 & 7.5173796(48) & 0.46121(23) & 69.23(11) & 47.505(92) & 2242.16105(74) & 8.979(19) & 1.810(28) \\ 
		\hline
		\multicolumn{8}{l}{The first column is the TESS ID; columns 2$-$8 are the 7 parameters in the K95$^+$ model.}\\
		\multicolumn{8}{l}{The unit of $T_{0p}$ is TJD=BJD$-$2,457,000.}\\
	\end{tabular}
\end{table*}

\begin{figure*}
	\includegraphics[width=0.75\textwidth]{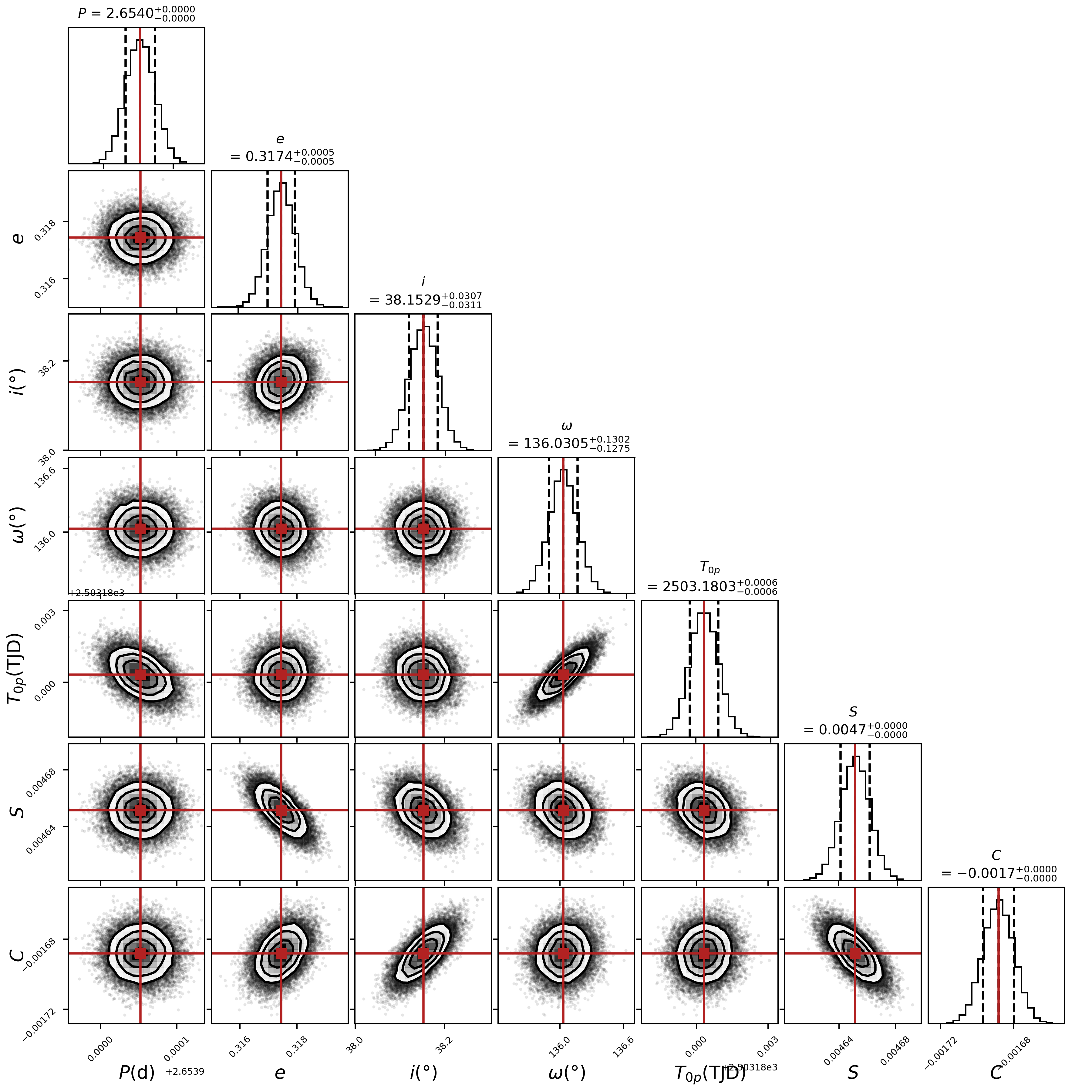}	
	\caption{Corner plot of the MCMC fit procedure for TIC 21222483. Red vertical lines indicate the median values of the presented histograms for each parameter. Black vertical dashed lines show 1 $\sigma$ uncertainties. The unit of $T_{0p}$ is TJD=BJD$-$2,457,000.
		\label{fig:corner}}
\end{figure*}

\begin{figure*}
	\subfloat[TIC 21222483]{\includegraphics[width=0.42\textwidth]{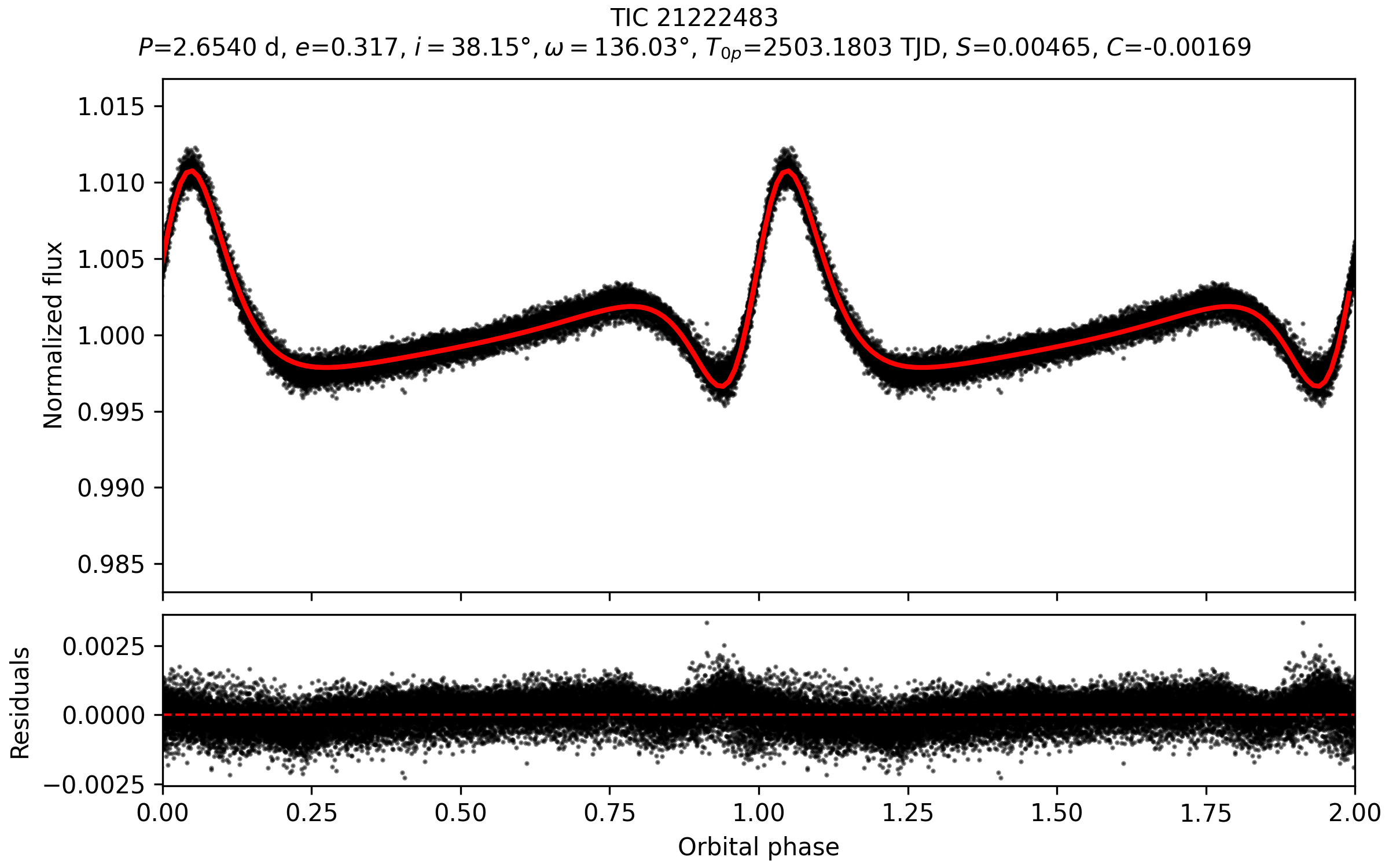}}
	\subfloat[TIC 58177975]{\includegraphics[width=0.42\textwidth]{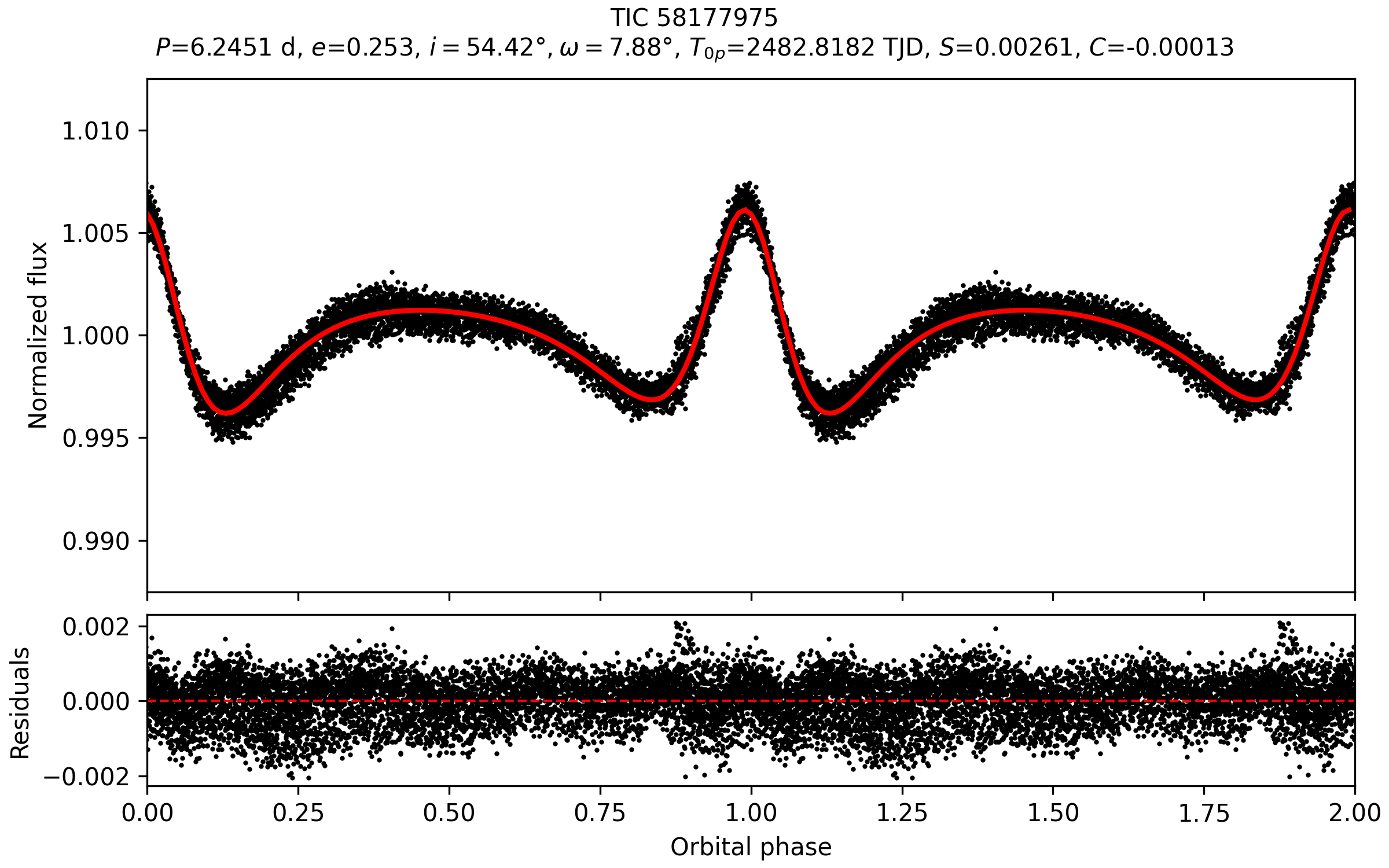}}\\
	\subfloat[TIC 71410794]{\includegraphics[width=0.42\textwidth]{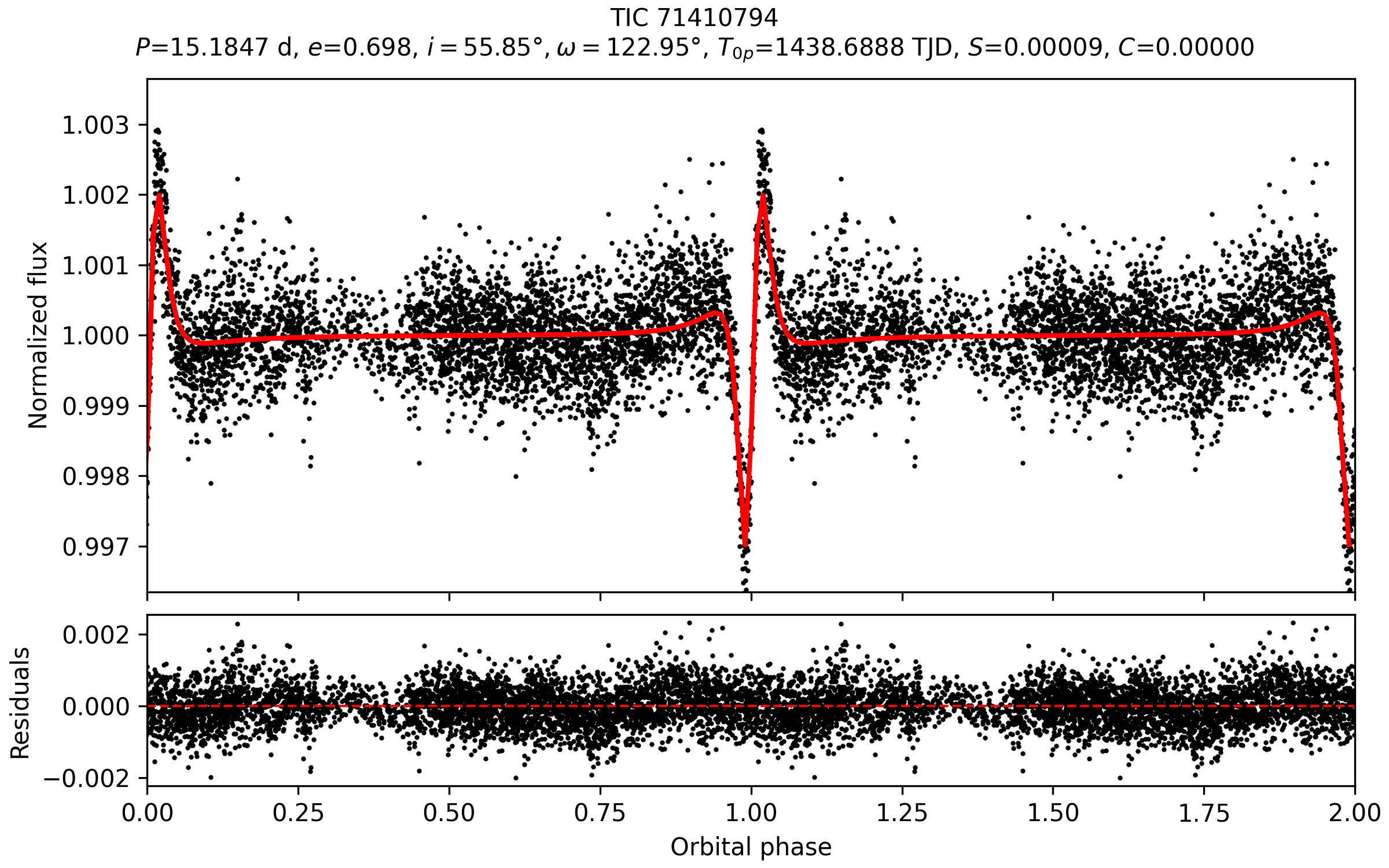}}
	\subfloat[TIC 85177749]{\includegraphics[width=0.42\textwidth]{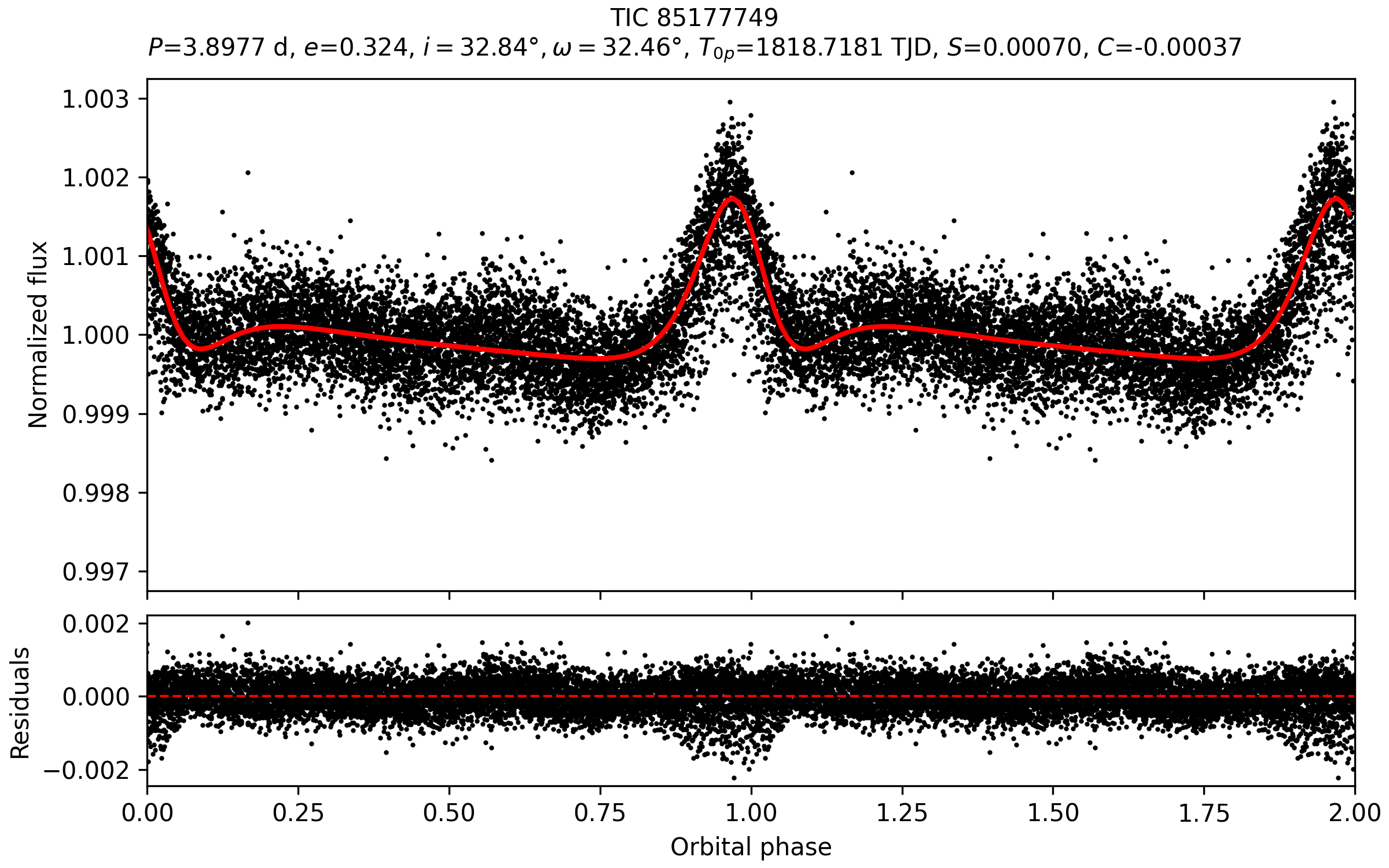}}\\
	\subfloat[TIC 118024242]{\includegraphics[width=0.42\textwidth]{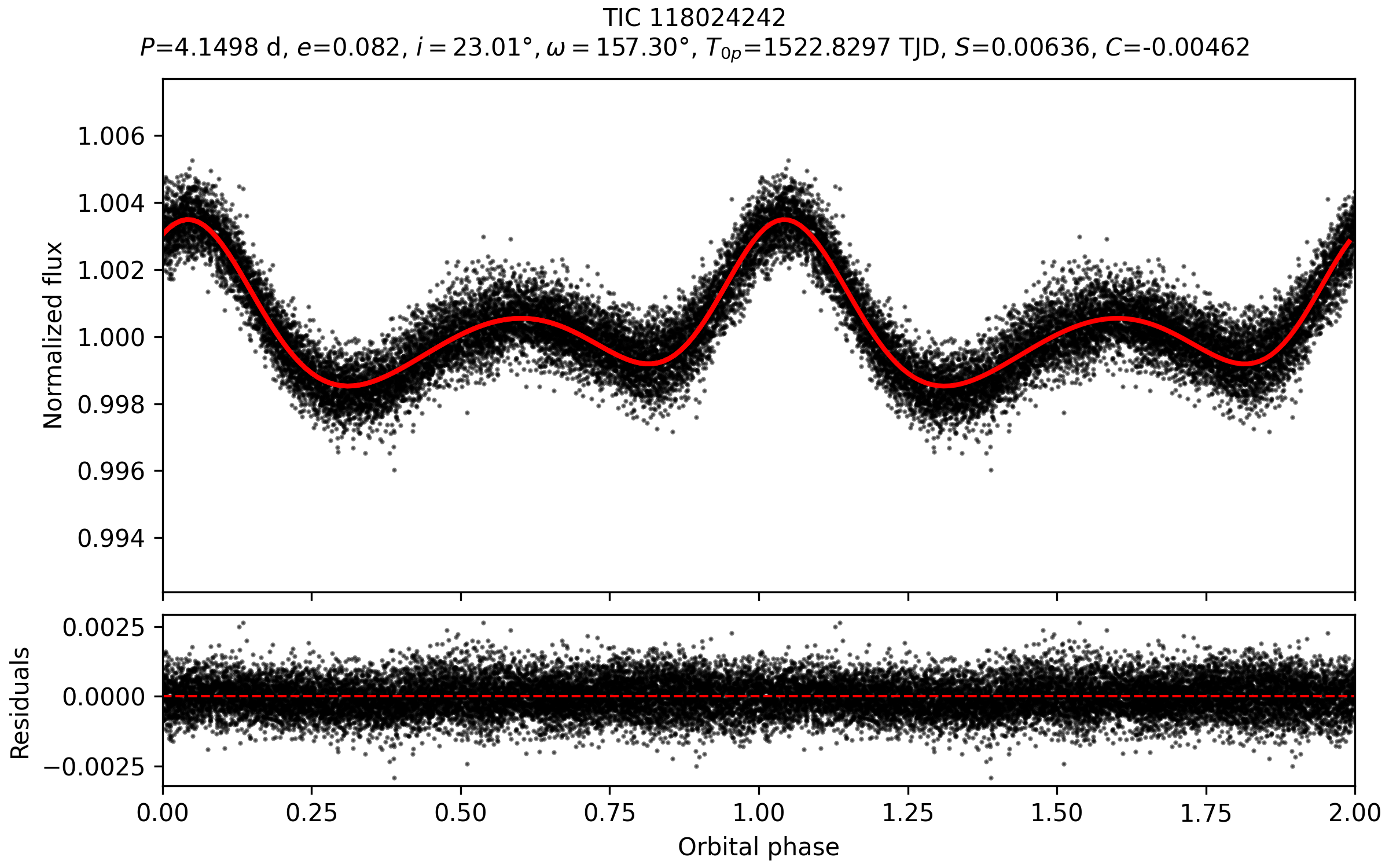}}
	\subfloat[TIC 137810570]{\includegraphics[width=0.42\textwidth]{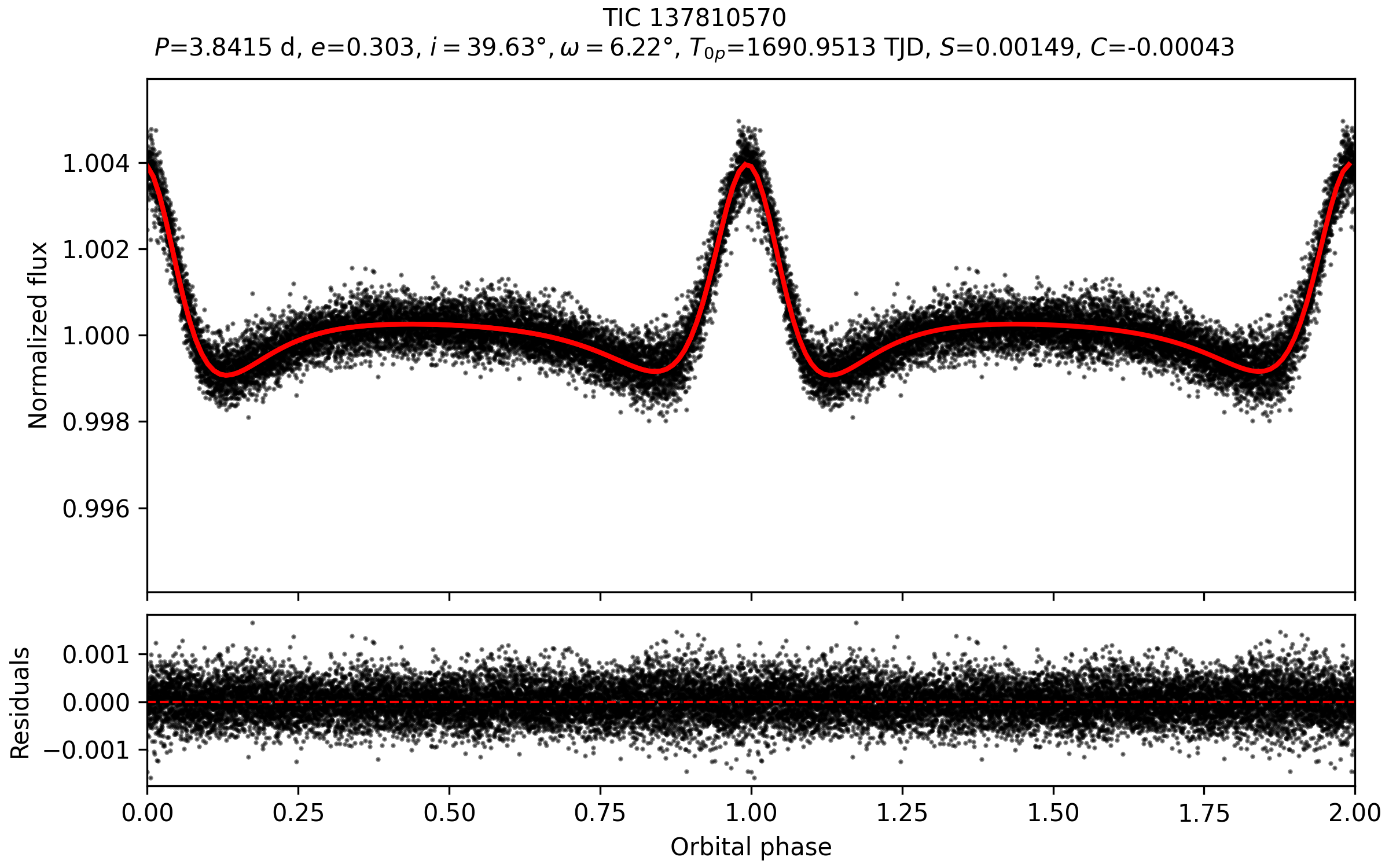}}\\
	\subfloat[TIC 153123772]{\includegraphics[width=0.42\textwidth]{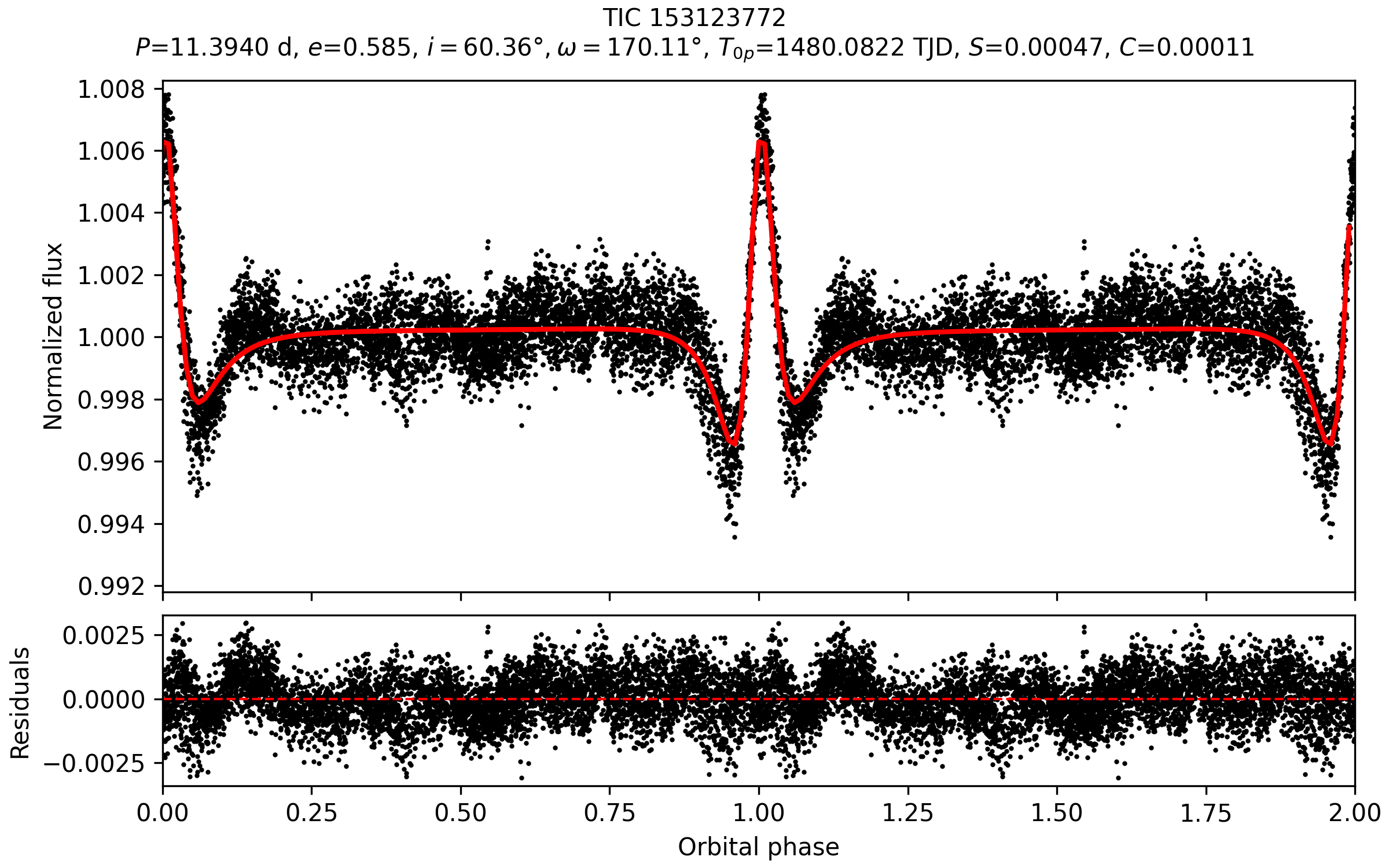}}
	\subfloat[TIC 153695072]{\includegraphics[width=0.42\textwidth]{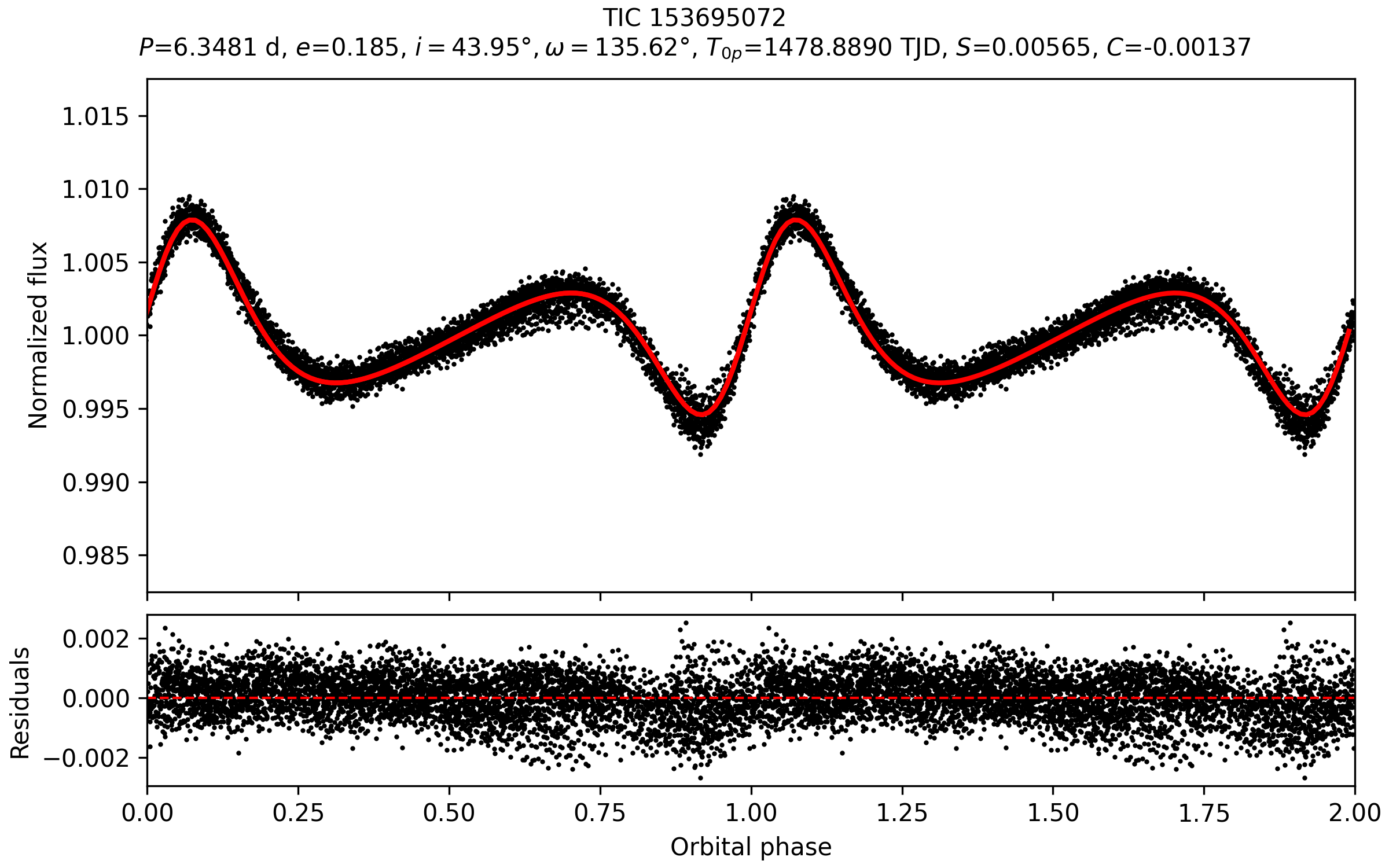}}\\
	
	\caption{The fit results of first eight HBSs. For each system, the K95$^+$ model (solid red line) fitted to the phase-folded light curve (black dots) in the upper panel;  the residuals of the fit are shown in the lower panel. The dashed red line indicates a zero point. Note: The plot shows phases 0$-$2 to make the fitting results clear, and phases 1$-$2 is an exact copy of phases 0$-$1.
		\label{fig:hbs1}}
\end{figure*}

\begin{figure*}
	\subfloat[TIC 202007047]{\includegraphics[width=0.42\textwidth]{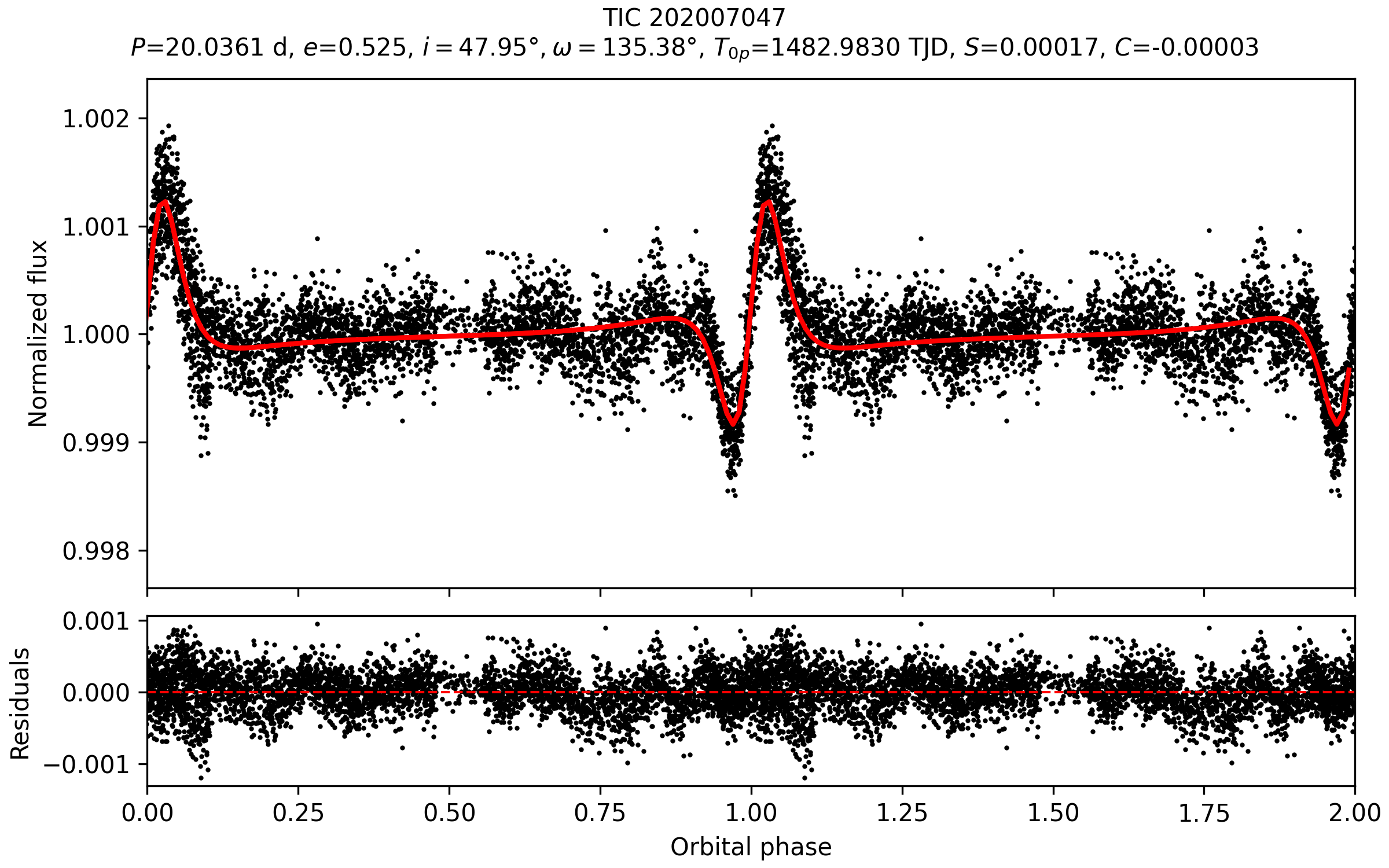}}
	\subfloat[TIC 262036561]{\includegraphics[width=0.42\textwidth]{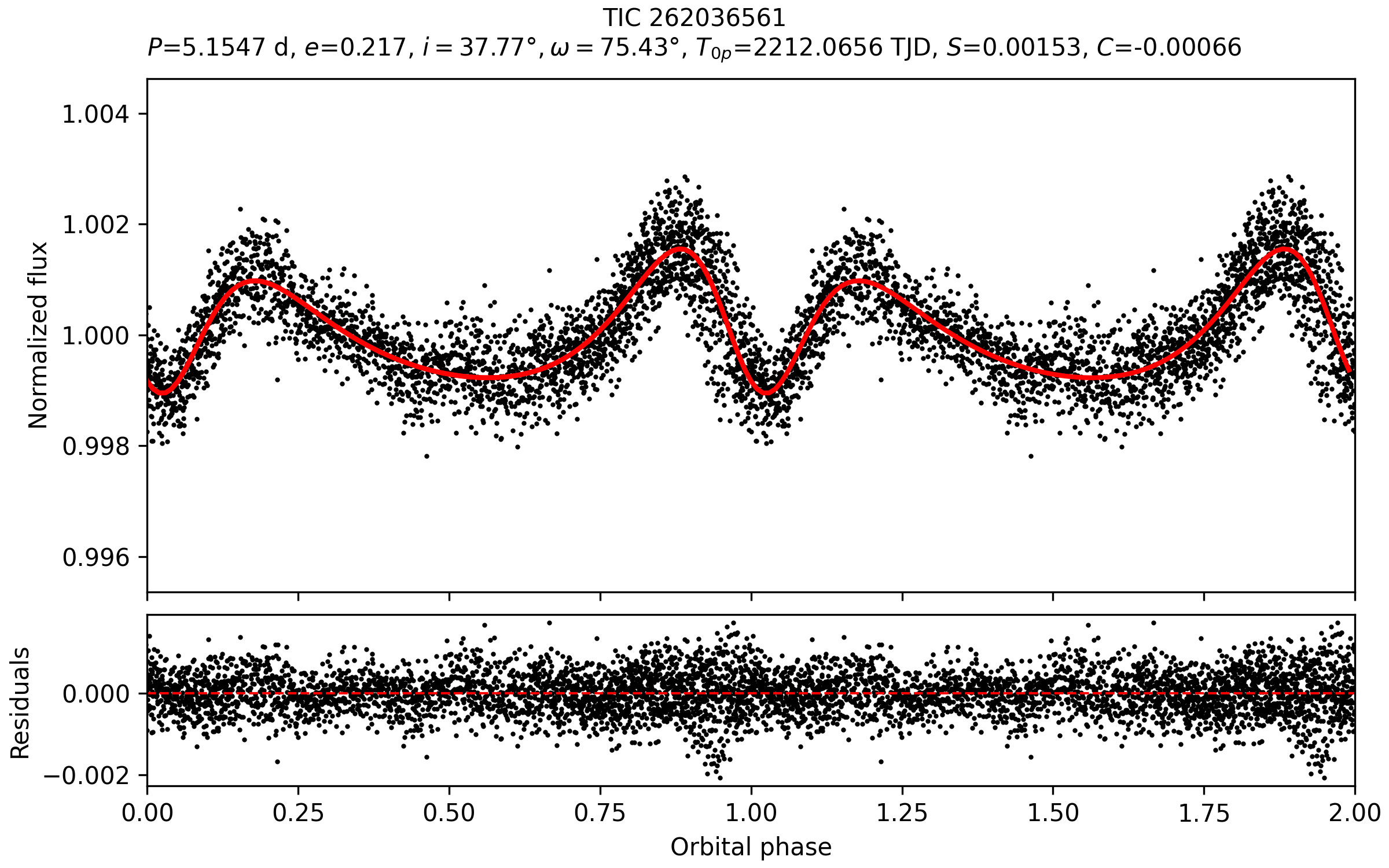}}\\
	\subfloat[TIC 265473090]{\includegraphics[width=0.42\textwidth]{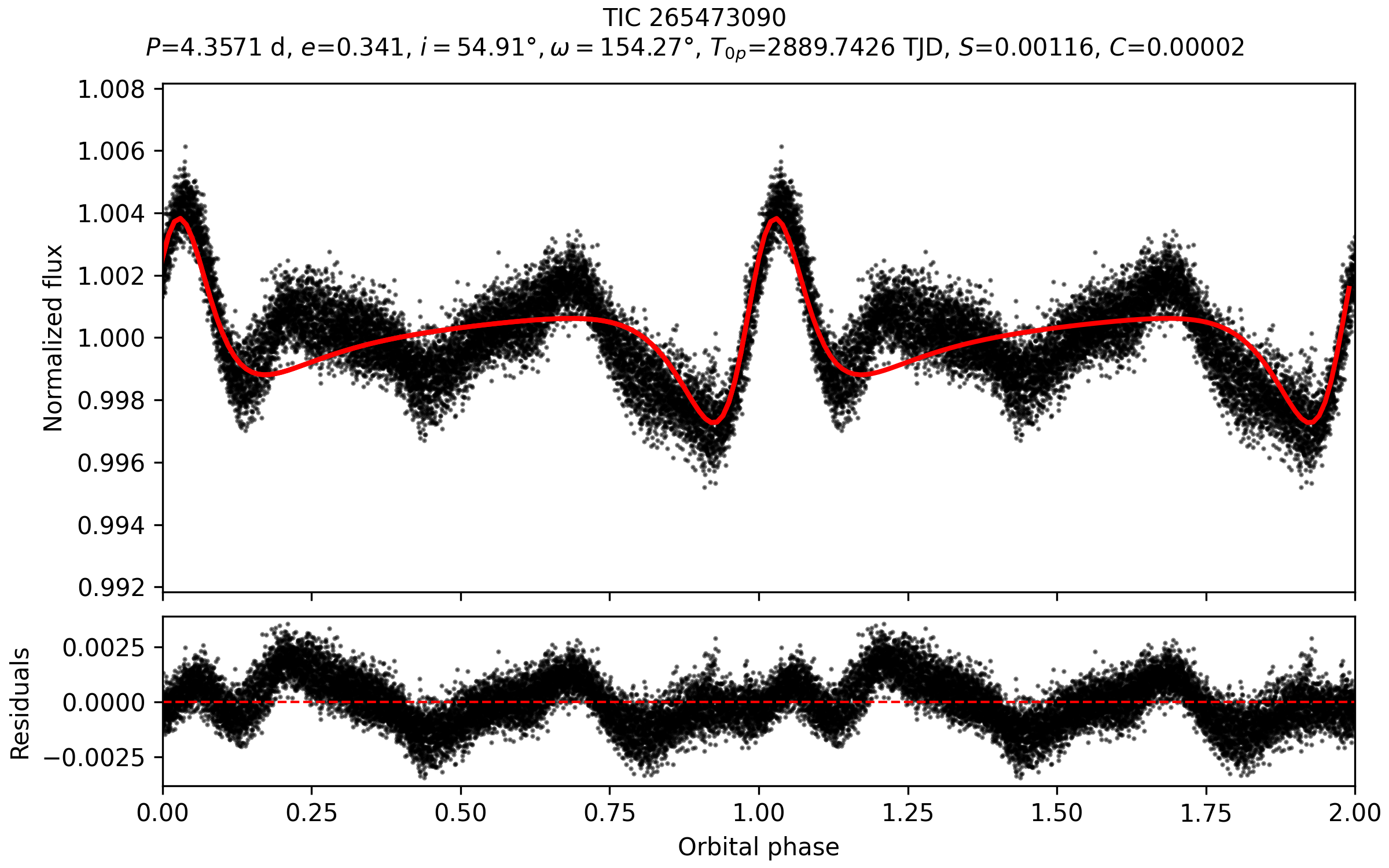}}
	\subfloat[TIC 283917392]{\includegraphics[width=0.42\textwidth]{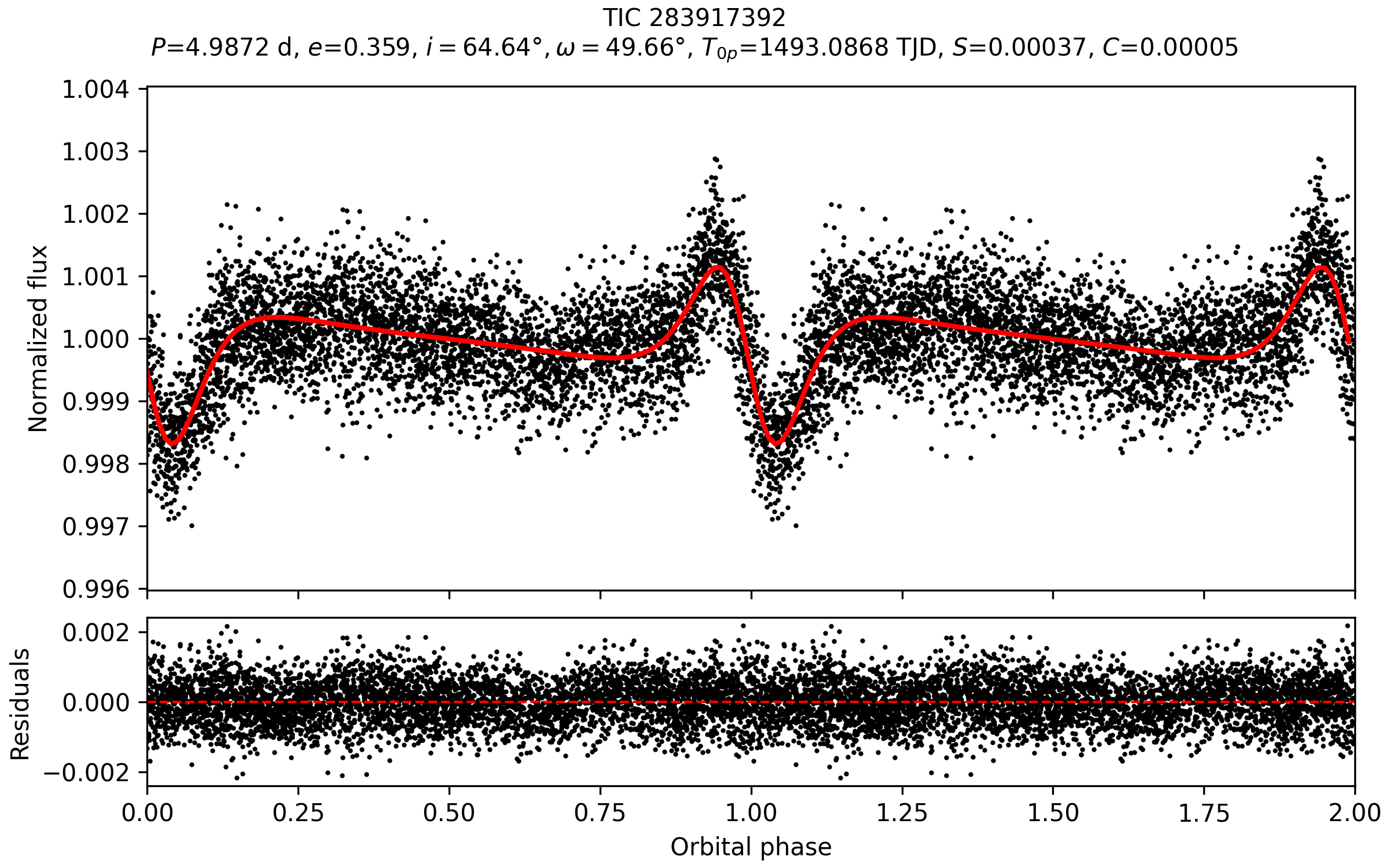}}\\
	\subfloat[TIC 292160826]{\includegraphics[width=0.42\textwidth]{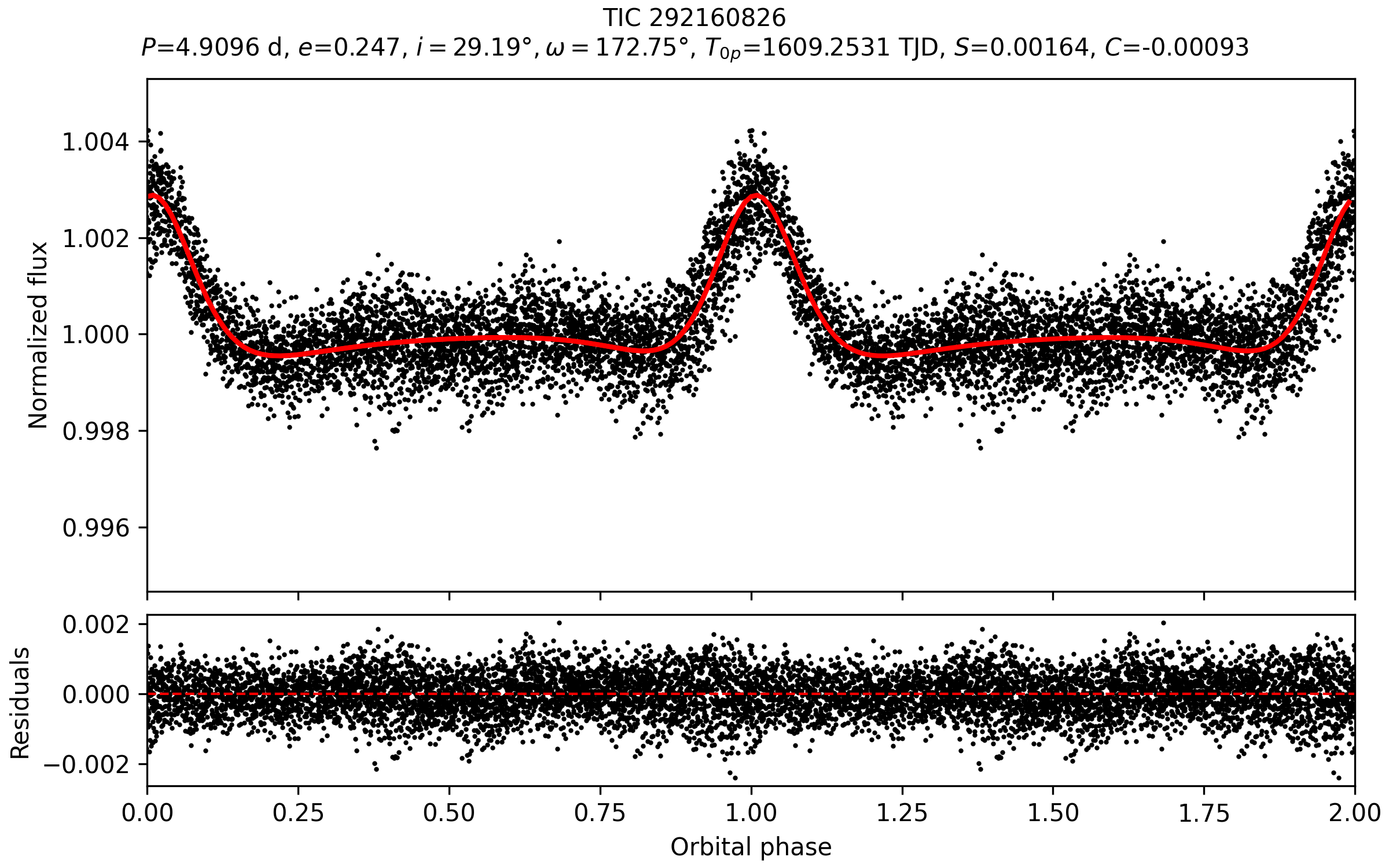}}
	\subfloat[TIC 353235026]{\includegraphics[width=0.42\textwidth]{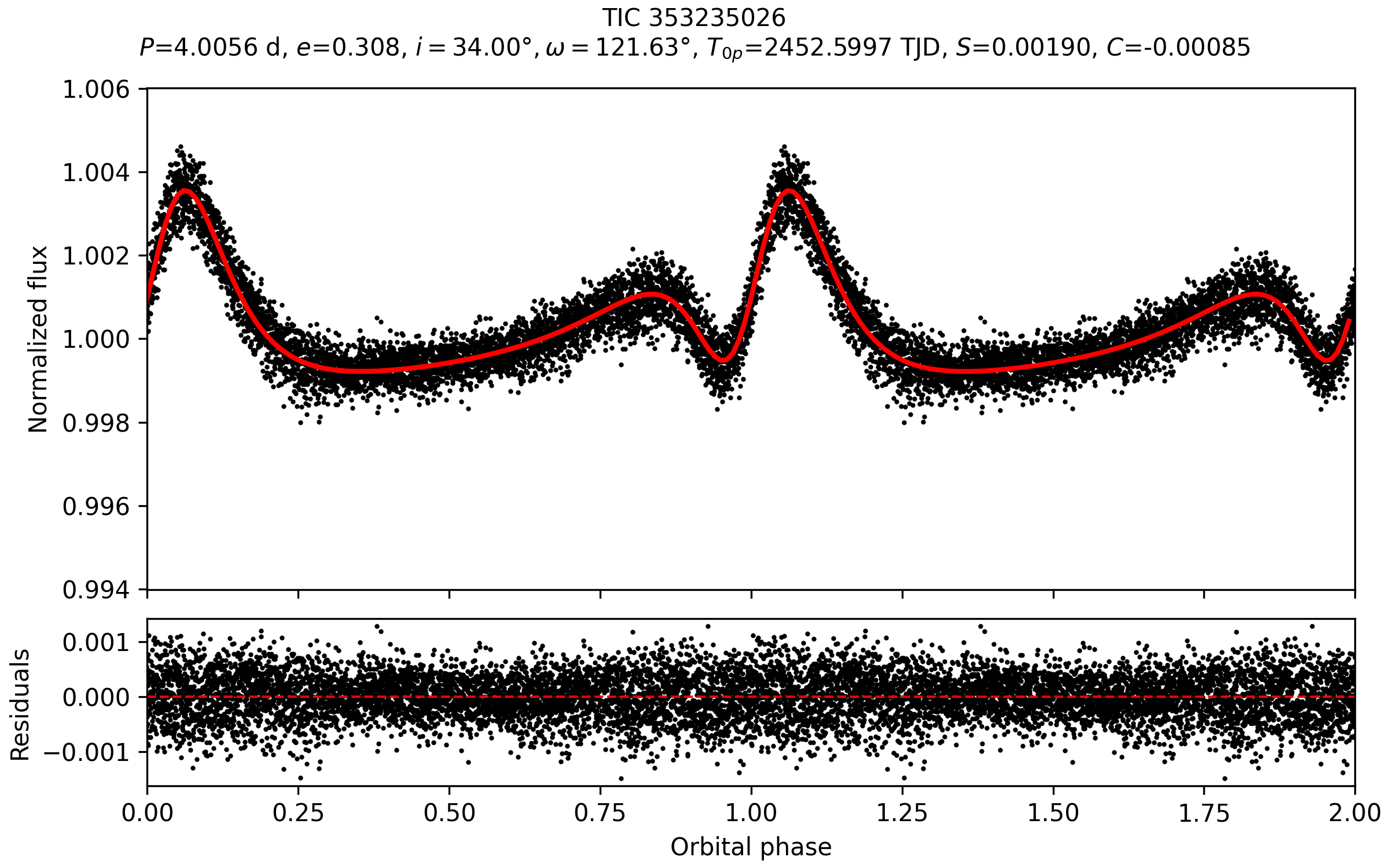}}\\
	\subfloat[TIC 363674490]{\includegraphics[width=0.42\textwidth]{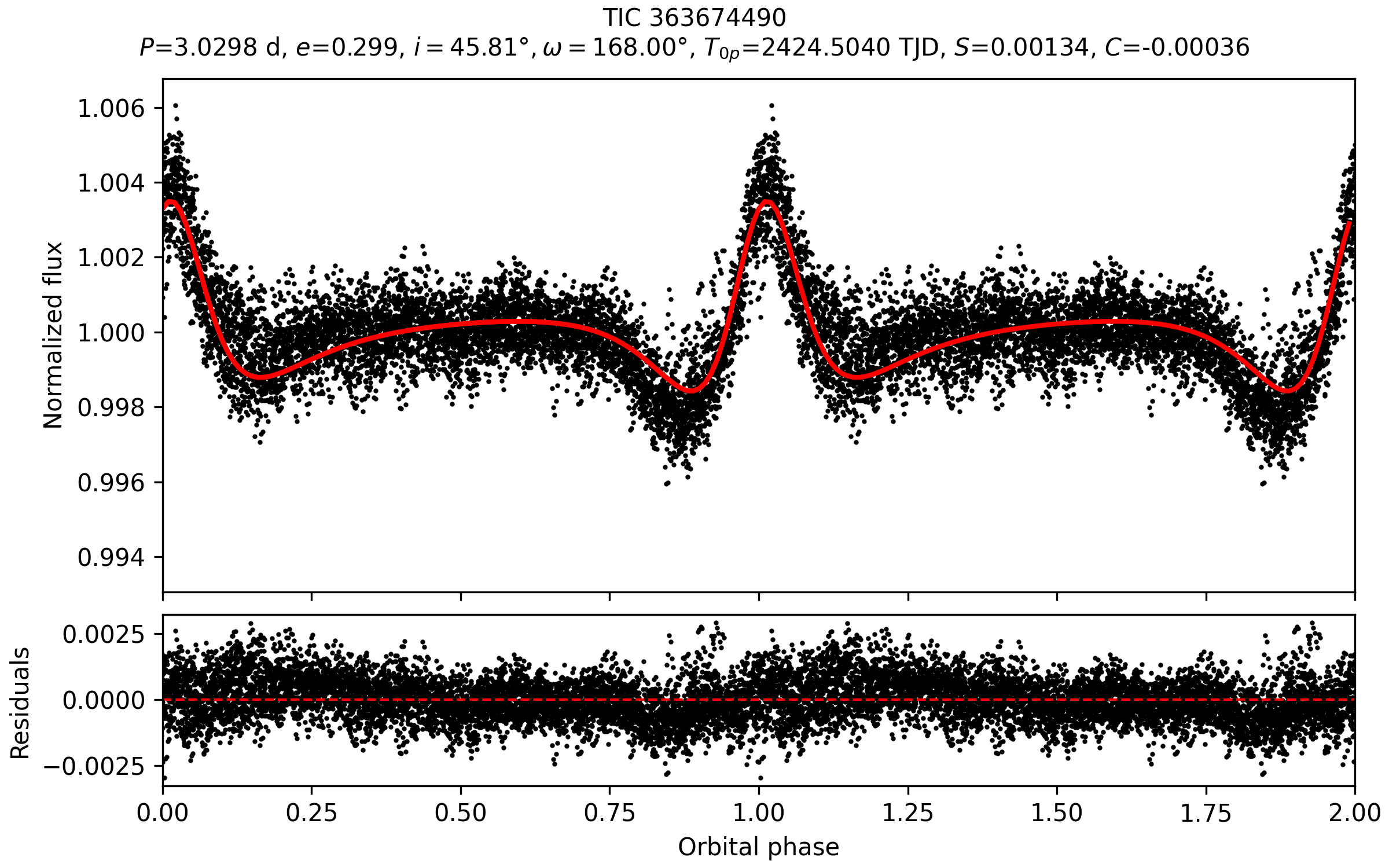}}
	\subfloat[TIC 370209445]{\includegraphics[width=0.42\textwidth]{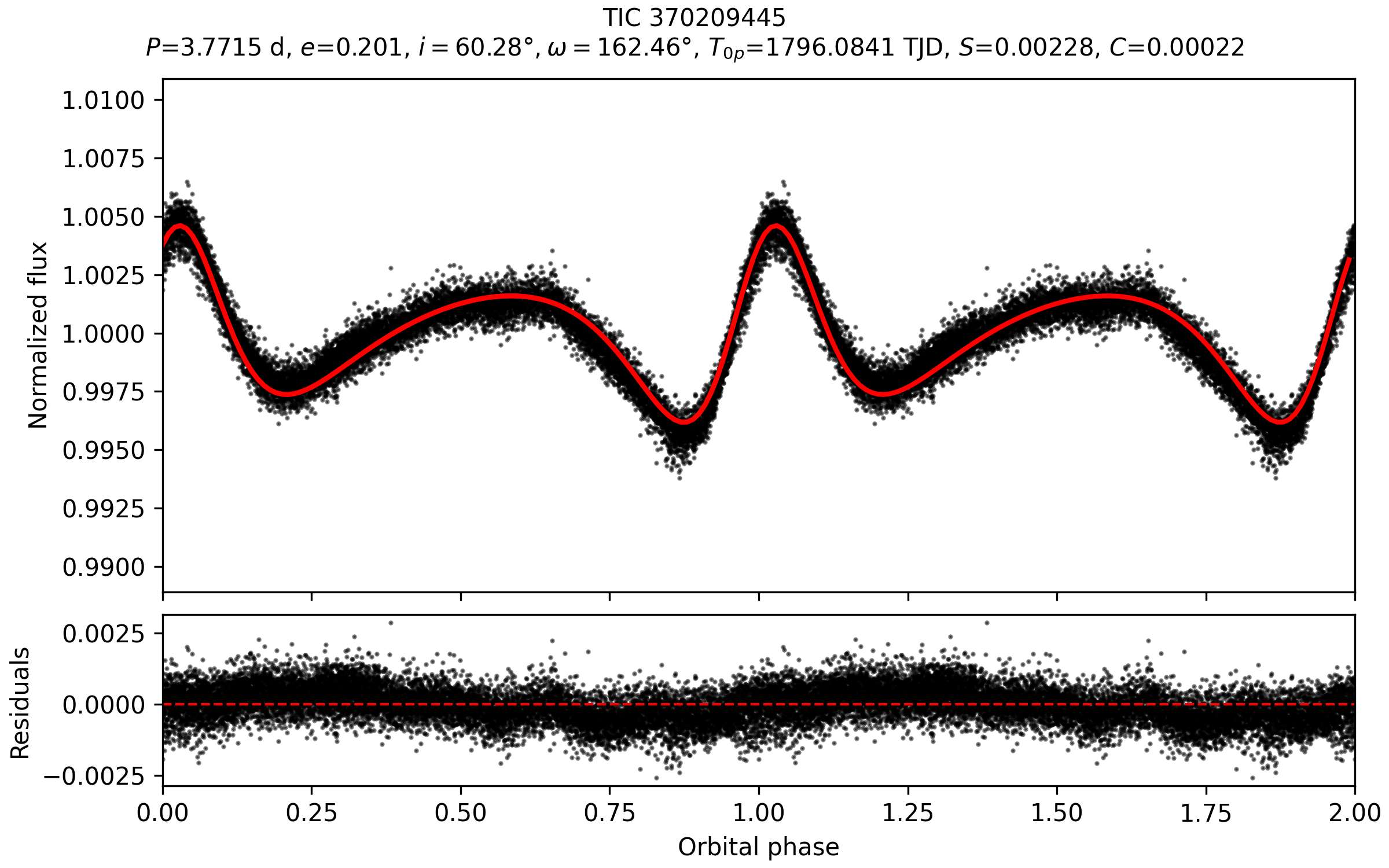}}\\
	\caption{Same as Figure \ref{fig:hbs1} for the other eight HBSs.
		\label{fig:hbs2}}
\end{figure*}

\begin{figure*}
	\subfloat[TIC 370269453]{\includegraphics[width=0.42\textwidth]{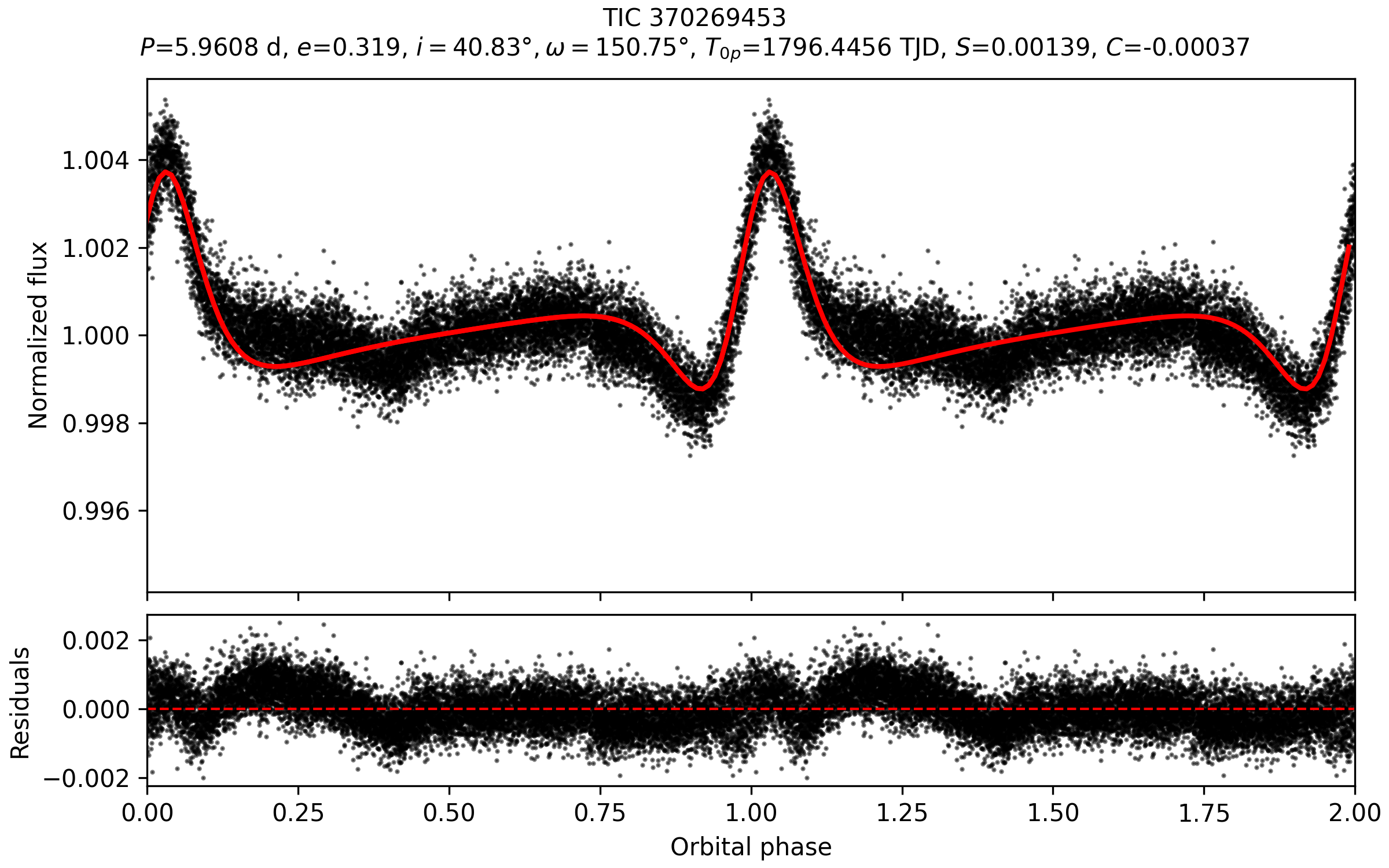}}
	\subfloat[TIC 386138719]{\includegraphics[width=0.42\textwidth]{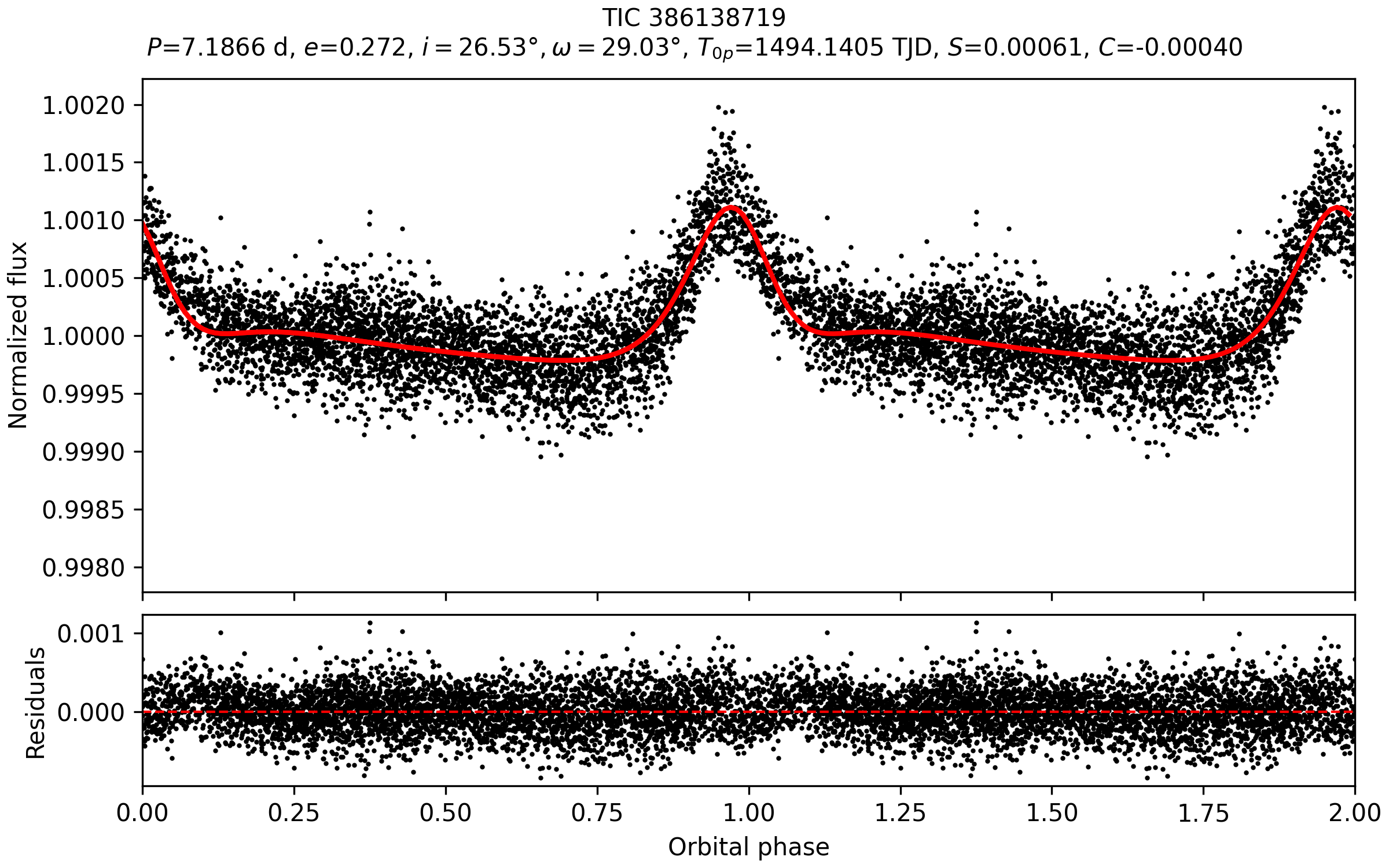}}\\
	\subfloat[TIC 388992242]{\includegraphics[width=0.42\textwidth]{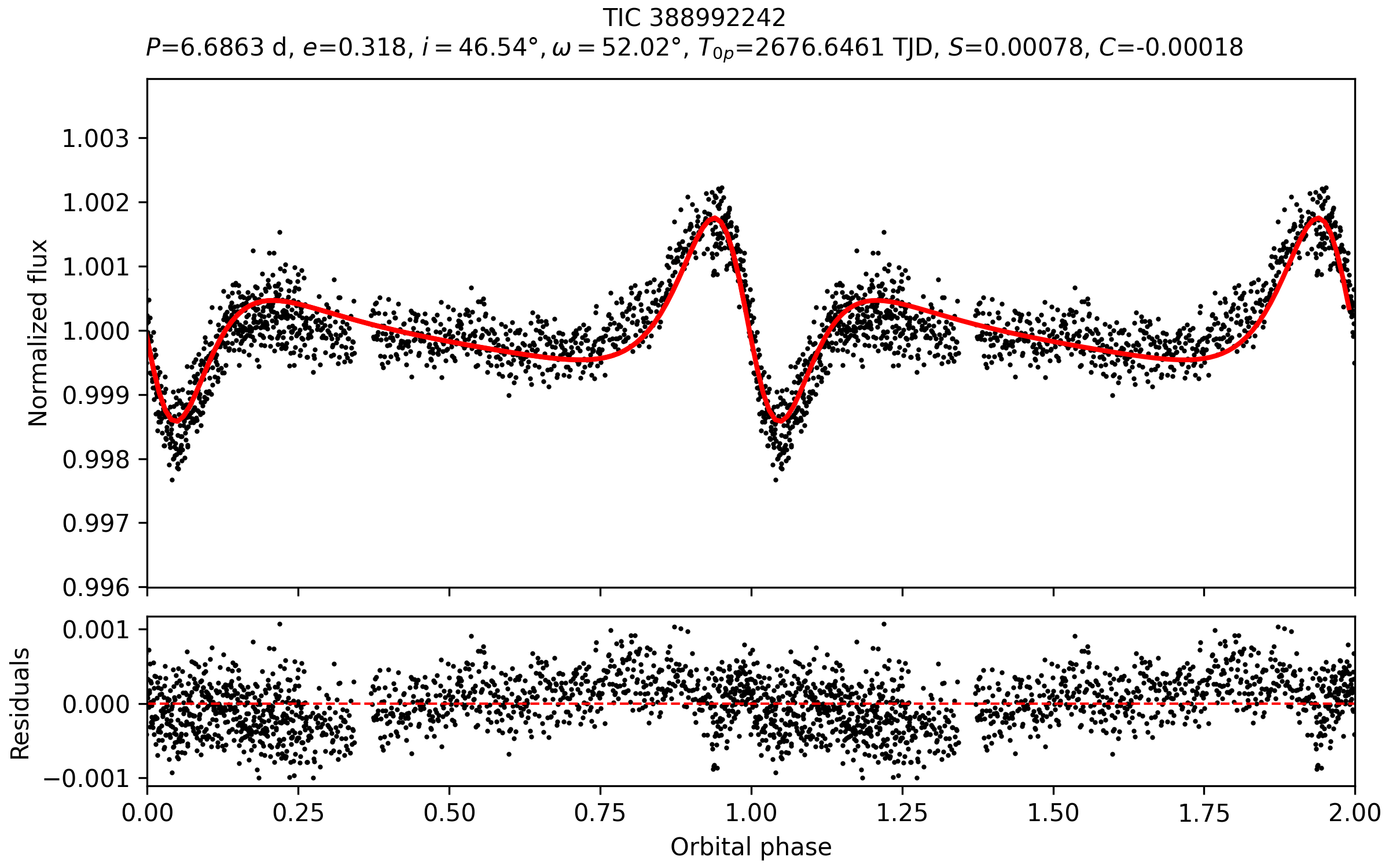}}
	\subfloat[TIC 411636838]{\includegraphics[width=0.42\textwidth]{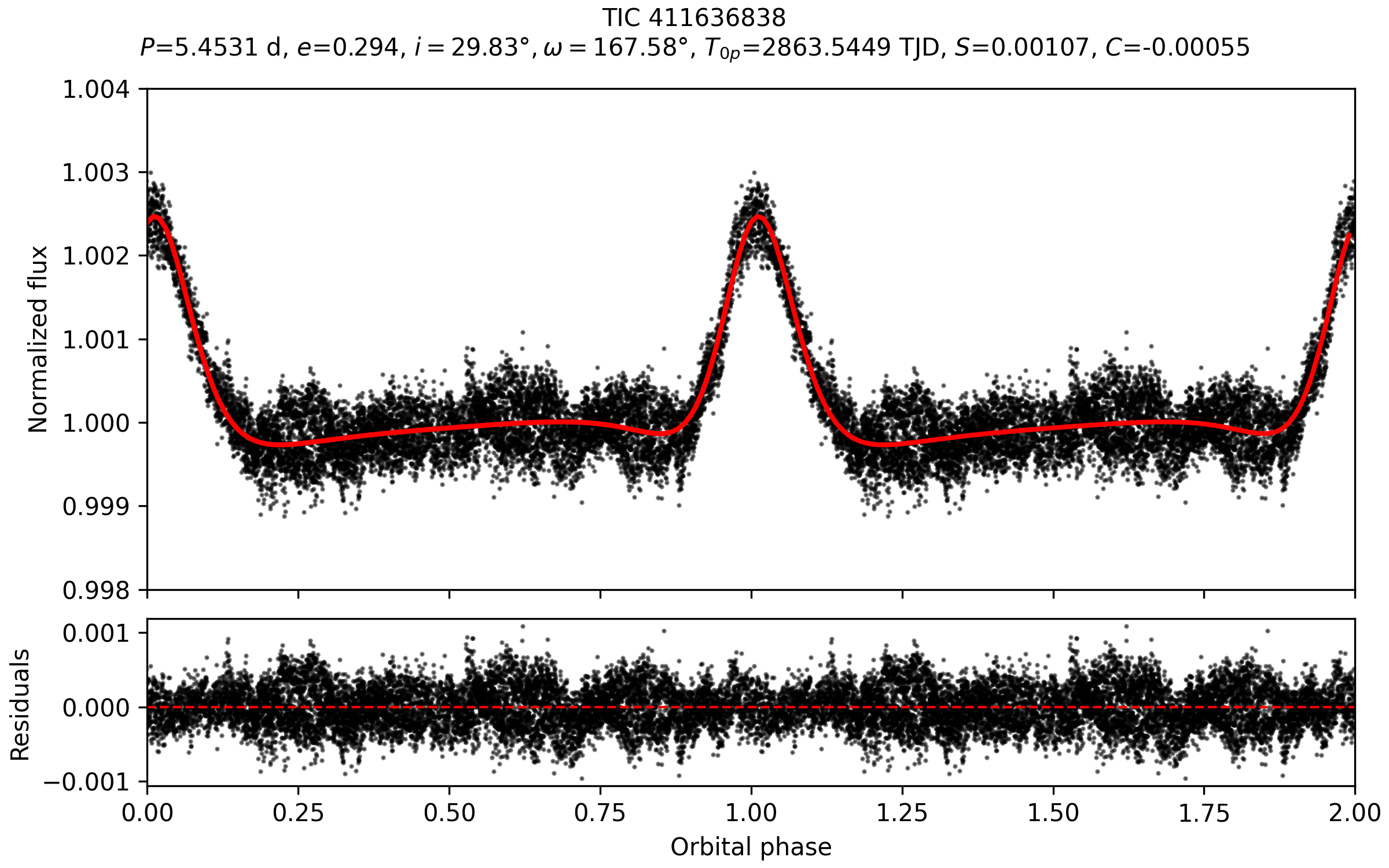}}\\
	\subfloat[TIC 412342944]{\includegraphics[width=0.42\textwidth]{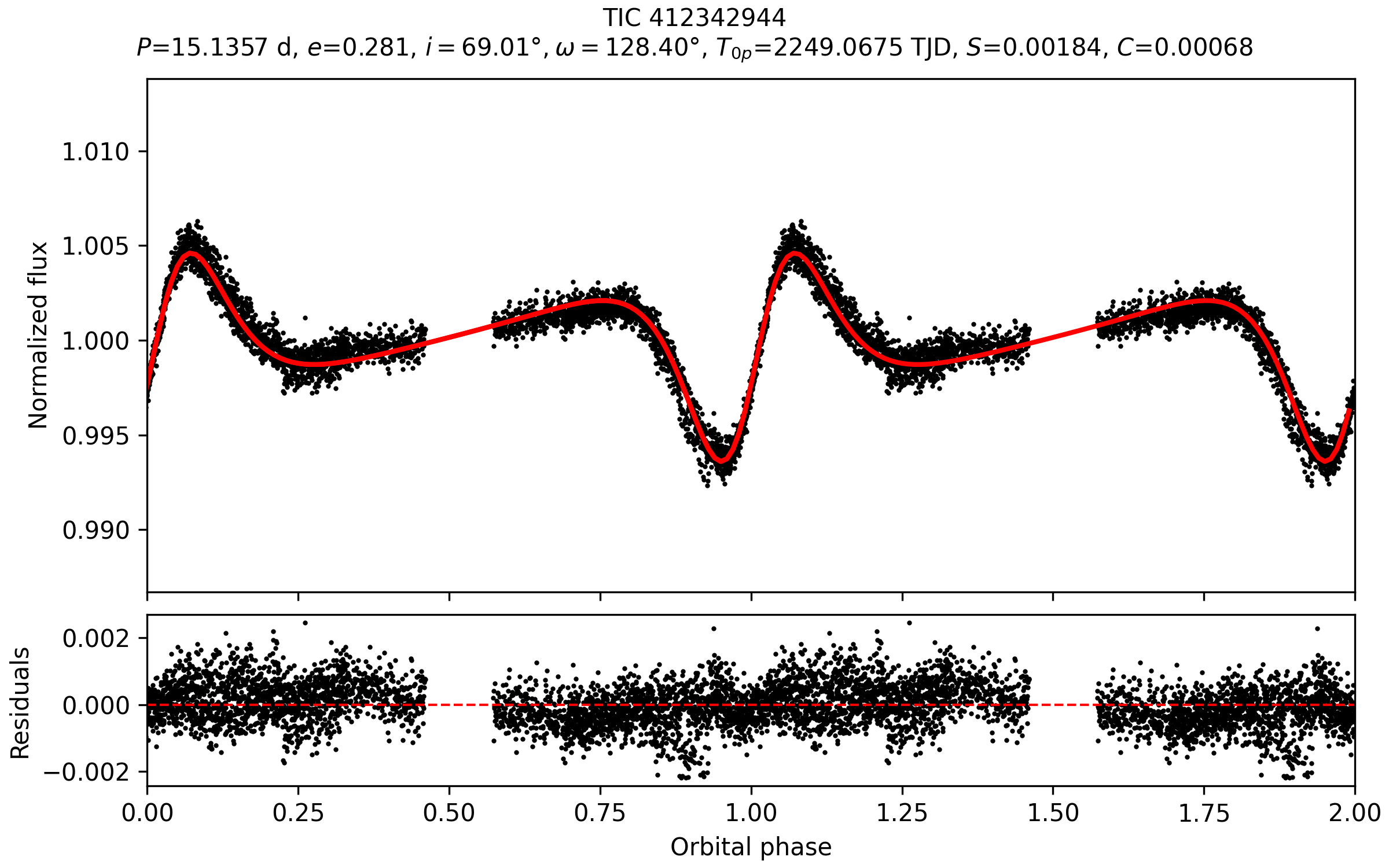}}
	\subfloat[TIC 444866141]{\includegraphics[width=0.42\textwidth]{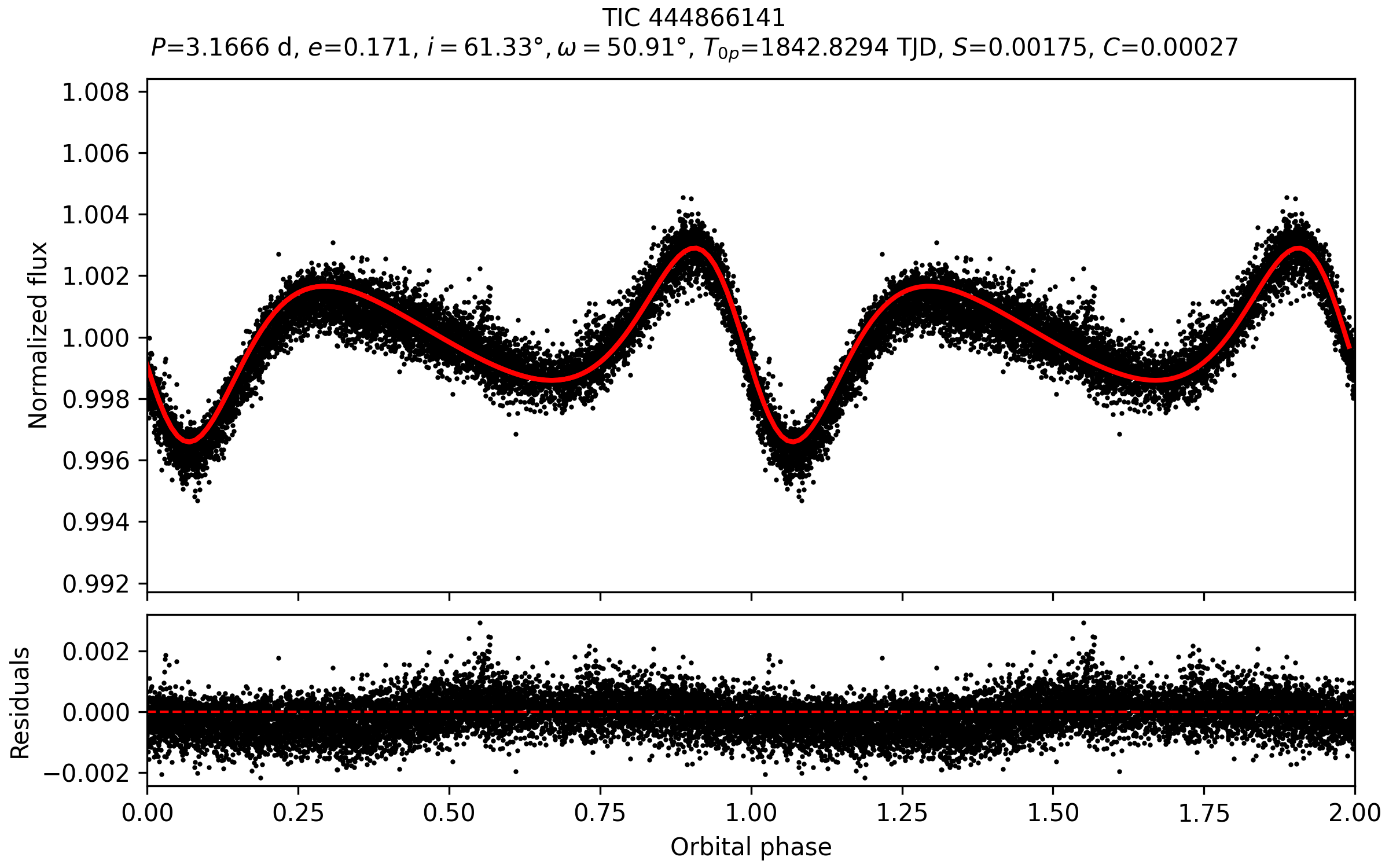}}\\
	\subfloat[TIC 452822117]{\includegraphics[width=0.42\textwidth]{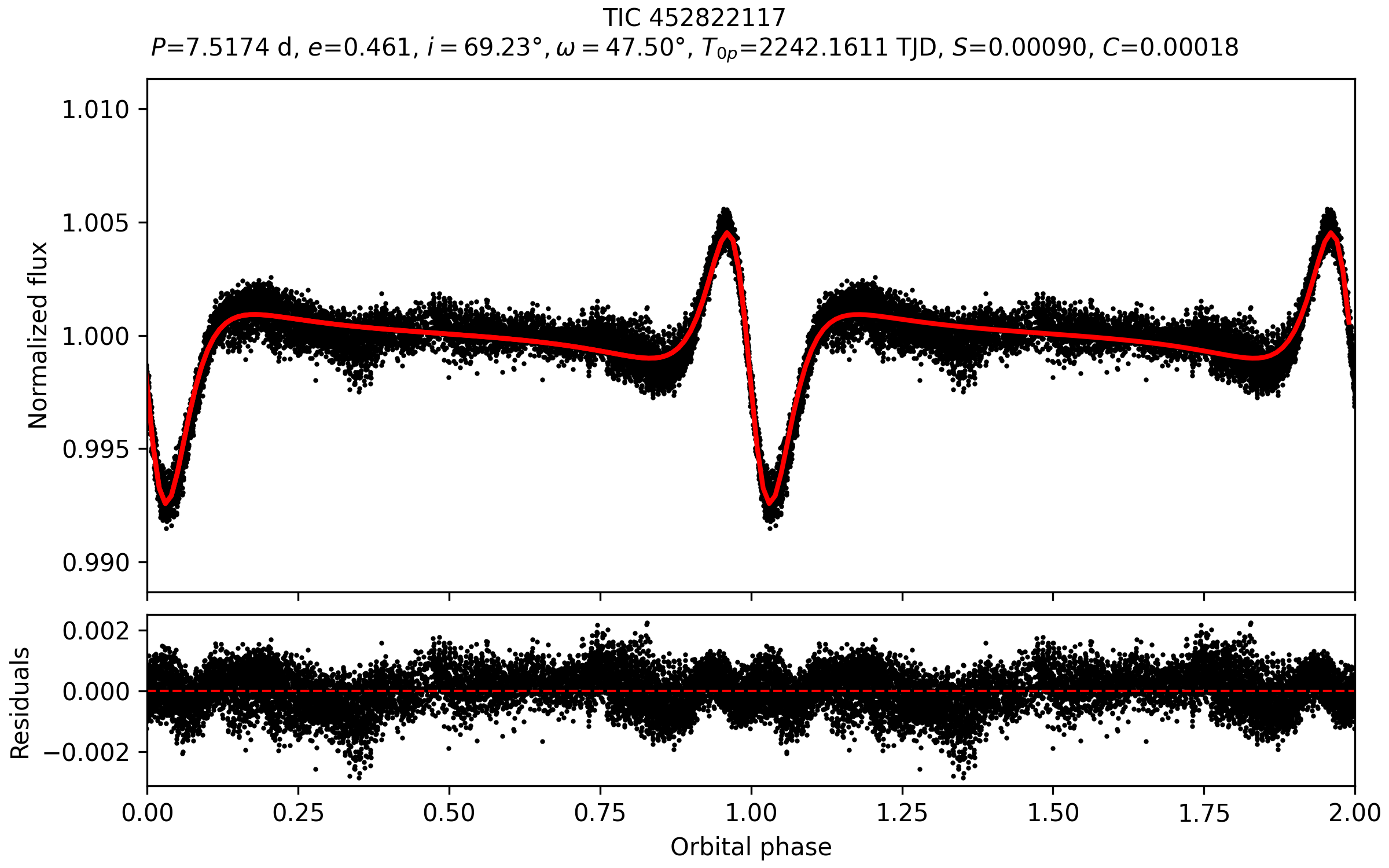}}
	\subfloat[All sectors of TIC 452822117]{\includegraphics[width=0.42\textwidth]{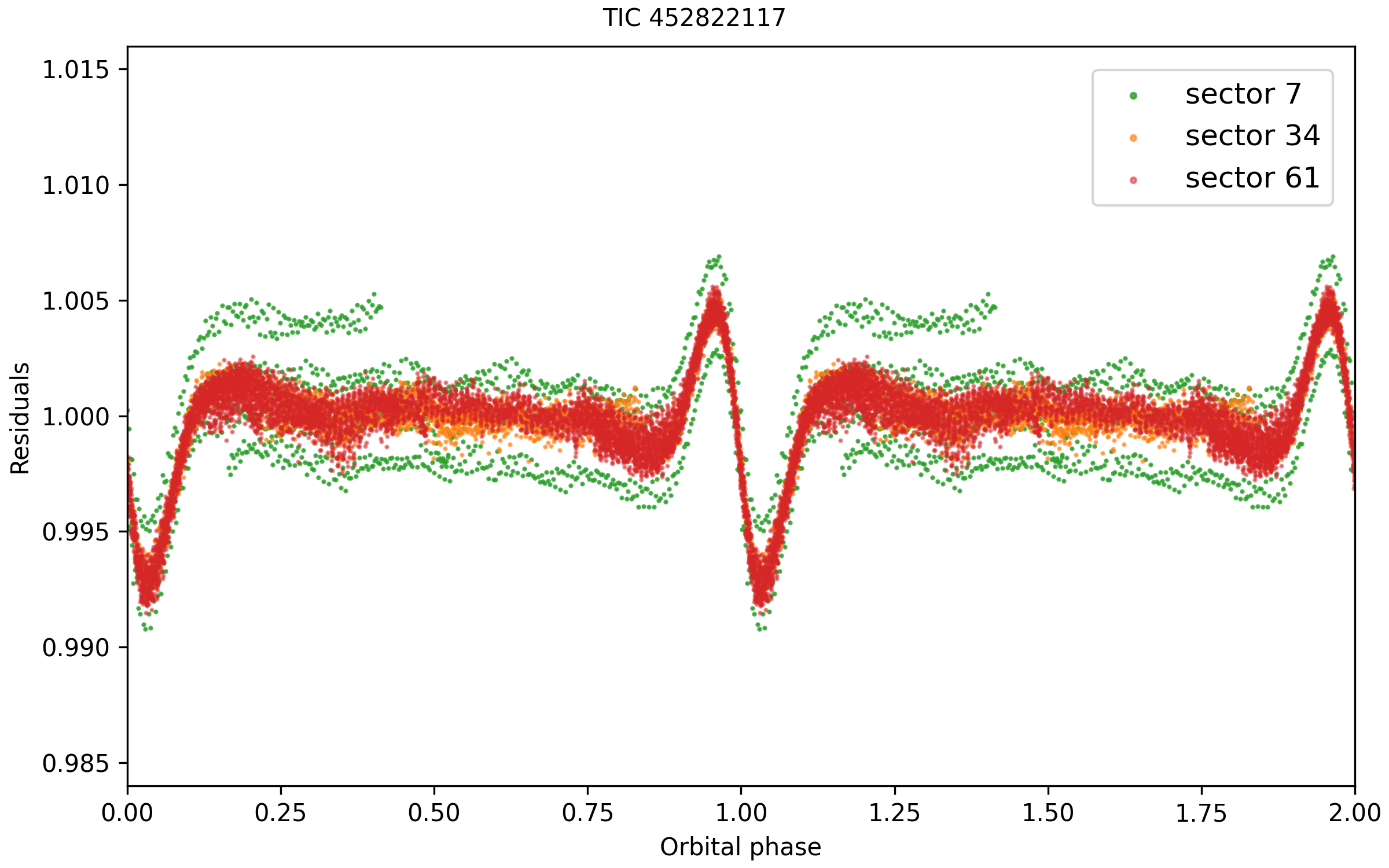}}\\
	\caption{Panels (a) $-$ (g) are the same as Figure \ref{fig:hbs1} for the other seven HBSs. Panel (h) shows the phase folded light curves of all three sectors of TIC 452822117 plotted in different colors.
		\label{fig:hbs3}}
\end{figure*}

\defcitealias{2021A&A...647A..12K}{KS21}
\defcitealias{2023ApJS..266...28L}{L23}

\section{The Properties of The New TESS HBSs} \label{sec:properties}
\subsection{The Eccentricity–Period Relation} \label{subsec:relation}
Figure \ref{fig:P_e} shows the eccentricity-period (e$-$P) diagram. The red stars represent the 23 TESS HBSs in this work. The green pluses indicate 20 TESS HBSs from (\citet{2021A&A...647A..12K}, hereafter \citetalias{2021A&A...647A..12K}). The gray circles are the Kepler HBSs reported in (\citet{2023ApJS..266...28L}, hereafter \citetalias{2023ApJS..266...28L}). The three black dashed curves represent the eccentricity-period relation of $e=\sqrt{1-(P_0/P)^{2/3}}$ \citep{2016ApJ...829...34S}, which is the expected functional form assuming conservation of angular momentum, with circularization periods $P_0$ of 3, 7, and 11 days. 

\begin{figure}
	\includegraphics[width=\columnwidth]{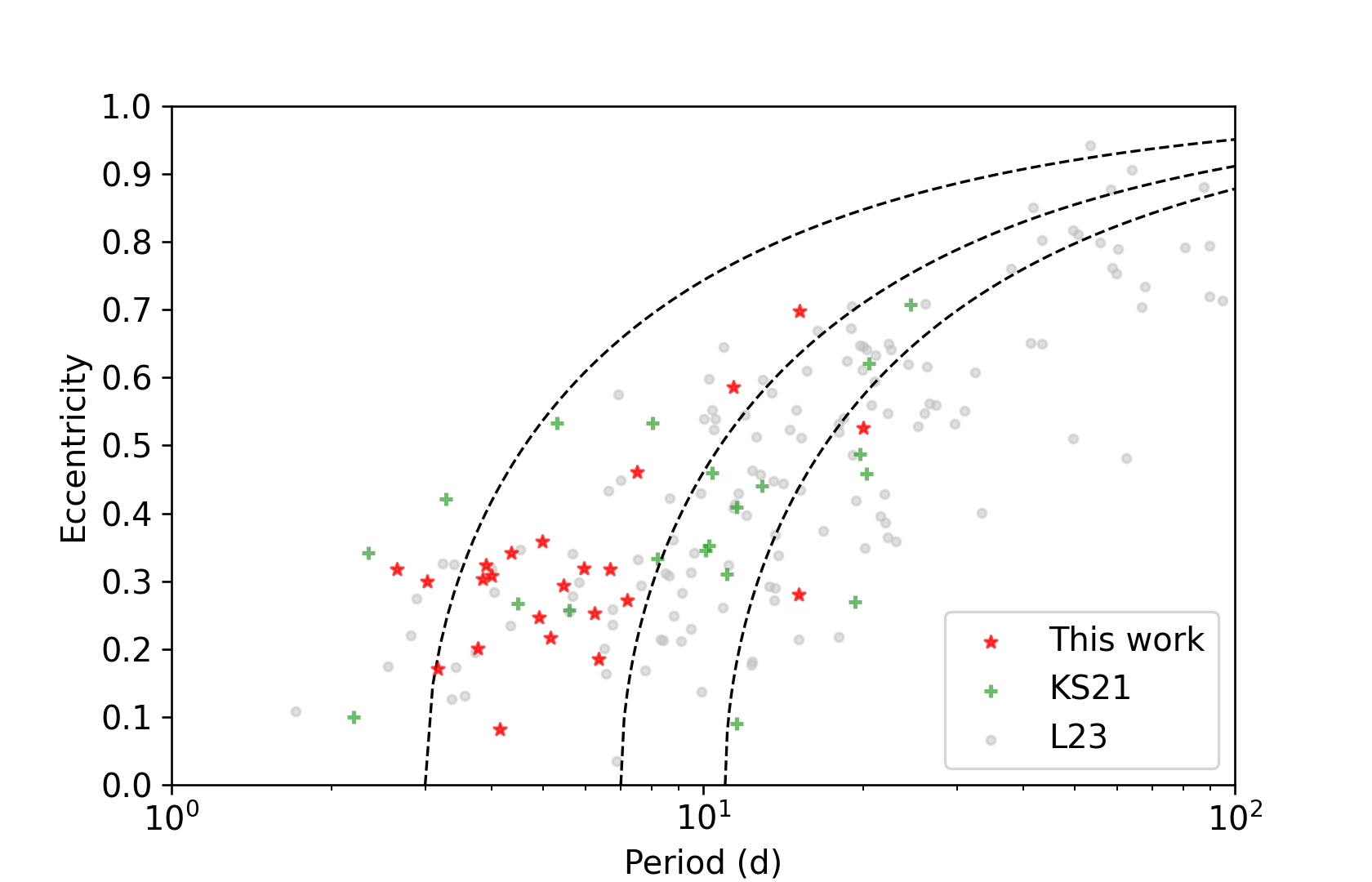}
	\caption{The eccentricity-period (e$-$P) diagram. Red stars indicate the TESS HBSs in this work. Green pluses indicate 20 TESS HBSs from \citetalias{2021A&A...647A..12K}. Gray circles represent the positions of the Kepler HBSs reported in \citetalias{2023ApJS..266...28L}. The three black dashed curves mark an eccentricity-period relation of $e=\sqrt{1-(P_0/P)^{2/3}}$, which is the expected functional form assuming conservation of angular momentum. The two curves use $P_0$ of 3, 7, and 11 days.
		\label{fig:P_e}}
\end{figure}

As can be seen, these TESS HBSs also do not exceed the upper envelope of the eccentricity-period distribution of the Kepler HBSs. Most of our samples are in the range of the first two black dashed curves, while the HBSs from \citetalias{2021A&A...647A..12K} are in a wider range (the first and the third dashed curves). This may indicate that these HBSs were born with a range of angular momenta and eventually tidally circularized to a range of periods $P_0$ in the range of 3 $-$ 11 days.

In addition, the e$-$P distribution of these objects also shows a positive correlation \citep{2023ApJS..266...28L}. It also shows the existence of orbital circularization \citep{2012ApJ...753...86T, 2016ApJ...829...34S} in HBSs. The shorter-period orbits are likely to circularize more quickly since they are undergoing stronger tidal forces; the eccentricity is smaller for shorter-period systems.

\subsection{The Hertzsprung-Russell Diagram of the HBSs}
\label{subsec:HR}
We obtain the parallax, visual magnitude, the interstellar extinction, and surface effective temperature of these HBSs from the Gaia Survey, which provides very high-precision astrometric data for nearly 2 billion stars \citep{2016A&A...595A...1G, 2018A&A...616A...1G, 2021A&A...649A...1G}, to illustrate the Hertzsprung-Russell (H-R) diagram. The luminosities of these systems are calculated using the following equations  \citep{2023ApJS..265...33S}:
\begin{equation}\label{equation:c}
	{\rm log}(L/L_{\odot})=0.4\cdot(4.74-M_V-BC)
\end{equation}
\begin{equation}\label{equation:d}
	M_V=m_V-5\cdot {\rm log}(1000/\pi)+5-A_V,
\end{equation}
where the bolometric correction BC is estimated following \citet{2023ApJS..265...33S} and \citet{2013ApJS..208....9P}; the visual magnitude $m_V$, the interstellar extinction $A_V$, and the parallax $\pi$ are from the Gaia Survey.

Figure \ref{fig:HR-all} shows the H-R diagram of the 23 HBSs in this work, TESS HBSs from \citetalias{2021A&A...647A..12K} and the Kepler HBSs from \citetalias{2023ApJS..266...28L}. The evolutionary tracks for the masses 1.5, 3, 4, and 7 M$_{\odot}$ from the theoretical zero-age main sequence (ZAMS) for Z = 0.02 are produced with the stellar evolution code Modules for Experiments in Stellar Astrophysics (MESA; \citet{2011ApJS..192....3P, 2013ApJS..208....4P, 2015ApJS..220...15P, 2018ApJS..234...34P, 2019ApJS..243...10P, 2023ApJS..265...15J}) version 22.05.1 and MESASDK 21.4.1 \citep{2021zndo...5802444T}. As can be seen, the HBSs are uniformly distributed. This indicates that the HBSs can appear in a much wider area of the H-R diagram. However, these TESS HBSs are located in the regions with higher surface temperatures and luminosities, which may be an observational effect. TESS is more likely to detect HBSs with higher mass.

\begin{figure}
	\includegraphics[width=\columnwidth]{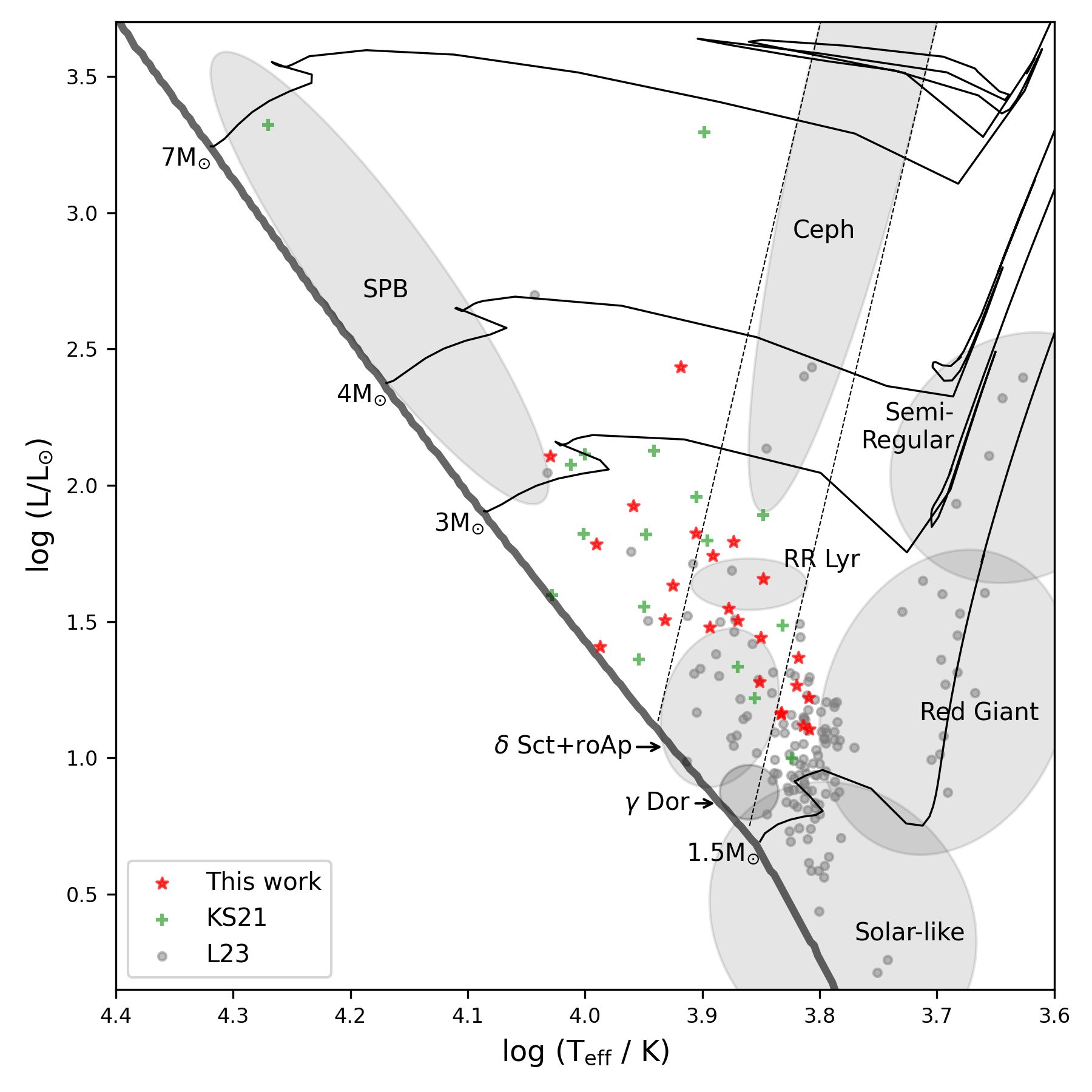}
	\caption{The H-R diagram of the HBSs. The red stars represent the 23 TESS HBSs in this work. The green pluses represent the HBSs from \citetalias{2021A&A...647A..12K}. The gray circles represent the Kepler HBSs from \citetalias{2023ApJS..266...28L}. The bold black solid curve represents the ZAMS. The black lines are the theoretical evolutionary tracks for the different masses with Z = 0.02. The gray dashed curves show the classical instability strip for radial pulsations. The classes of pulsating variables, including SPB, Ceph, RR Lyr, $\delta$ Sct+roAp, $\gamma$ Dor, semi-regular, red giant, and solar-like, are labeled next to the corresponding gray regions, which are derived according to \citet{2013PhDT.........6P, 2019ApJS..243...10P, 2021RvMP...93a5001A}.
		\label{fig:HR-all}}
\end{figure}

\subsection{Distribution of the Parameters}
Figure \ref{fig:dis} presents histograms of the orbital parameters $P$, $e$, $i$, $\omega$,  effective temperature, and luminosity from \citetalias{2023ApJS..266...28L}, \citetalias{2021A&A...647A..12K}, and this work in different colors. In each panel, the black curves represent the HBSs from \citetalias{2023ApJS..266...28L}; the green curves represent the HBSs from \citetalias{2021A&A...647A..12K}; the red curves represent the 23 HBSs in this work. In addition, the yellow areas represent all TESS HBSs (this work and \citetalias{2021A&A...647A..12K}). In panel (a), all objects have periods smaller than 20 days, most of them less than 10 days. We assume that this is due to observational effects. Since each sector of the TESS survey is continuously observed for 27 days, it is difficult to detect long-period HBSs.

In panel (b), the distribution of eccentricities is mostly less than 0.4, and only a few systems have higher eccentricities. This is again due to observational effects. Because the higher eccentricity systems tend to have longer periods that do not exist. 

In panels (c) and (d), the overall distributions of $i$ and $\omega$ are nearly uniform, but in the absence of $i>$ 70$^\circ$, $i<$ 20$^\circ$ and $\omega$ close to 90$^\circ$. On the one hand, the low inclination HBSs may mimic other types of variability, such as spotted stars or some kind of Be stars. On the other hand, for HBSs with high inclination and $\omega$ near 90$^\circ$, their light curves are similar to the eclipsing binaries or the classical ellipsoidal variables \citep{2022ApJ...928..135W}. Therefore, the absence of $i$ and $\omega$ may be due to selection effects.

In panels (e) and (f), the distributions show that these HBSs have higher temperatures ($T_{\rm eff}$ $\gtrsim$ 7000 K) and luminosities (${\rm log}(L/L_{\odot})$ $\gtrsim$ 1.1). Our findings suggest that the TESS survey might be more effective at detecting massive HBSs. Further investigation is warranted to quantify this potential bias. In addition, this may also be due to selection effects, as the brighter TESS samples may be selected in the works.

\begin{figure*}
	\subfloat[]{\includegraphics[width=0.5\textwidth]{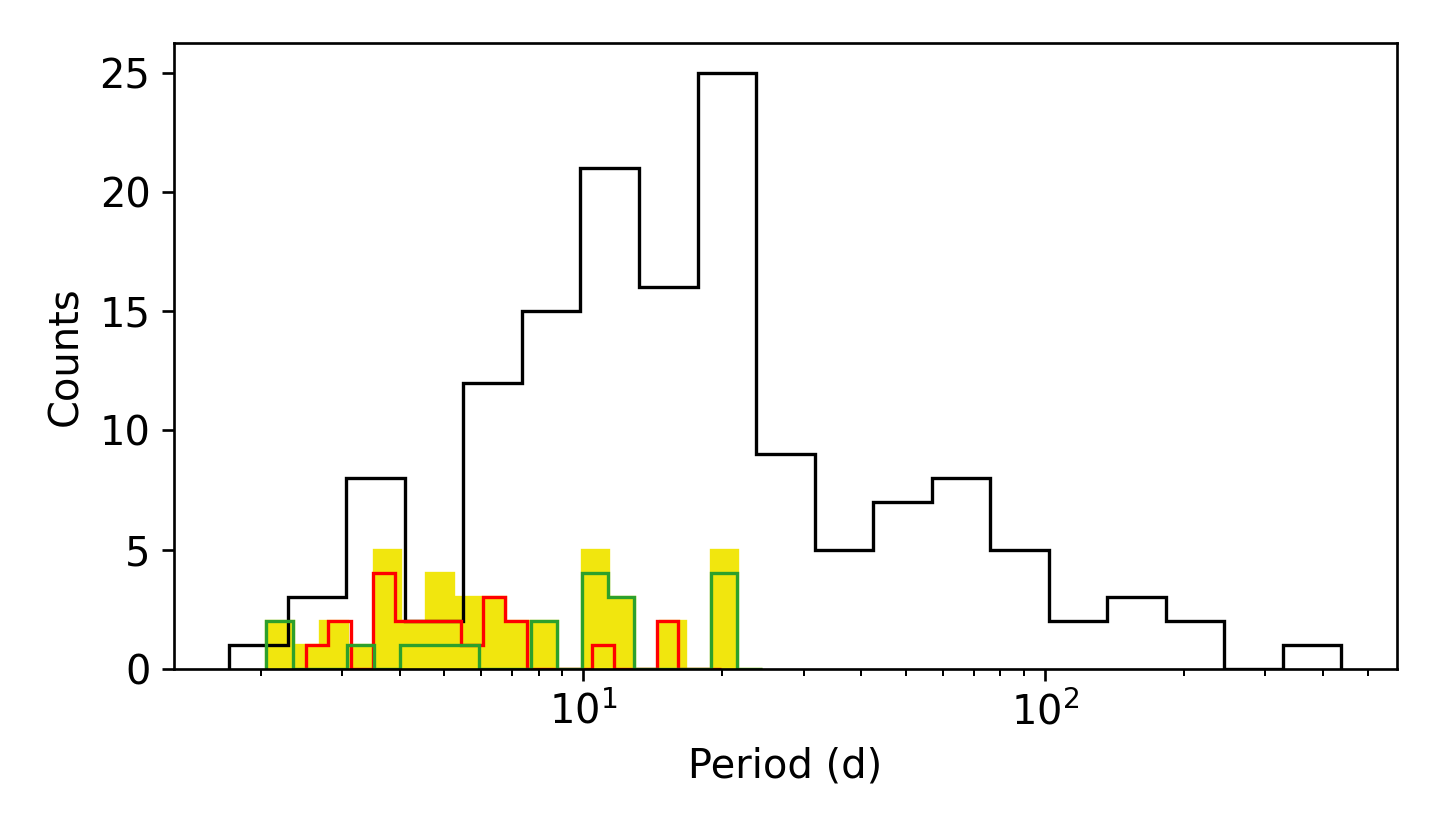}}
	\subfloat[]{\includegraphics[width=0.5\textwidth]{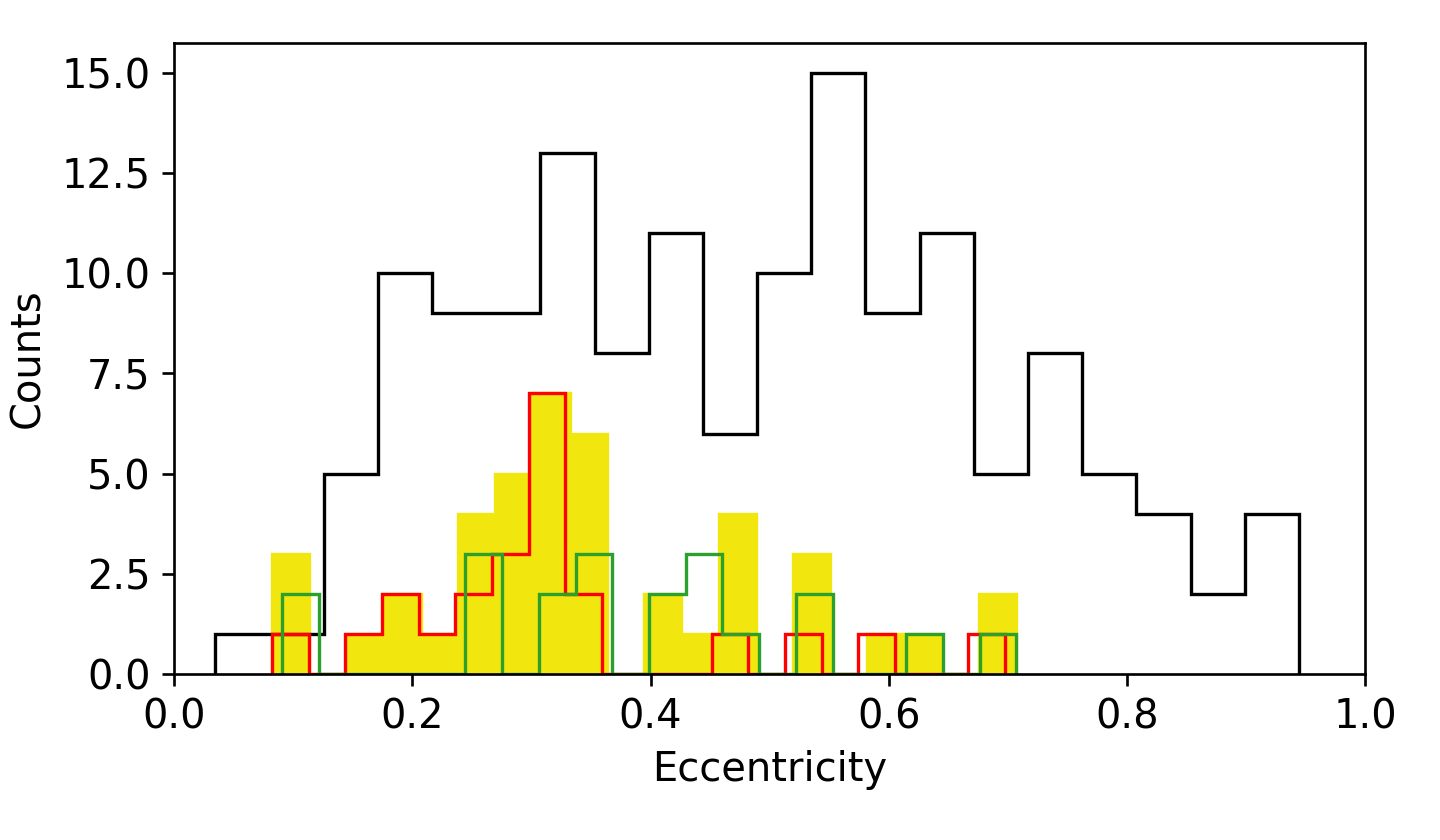}}\\
	\subfloat[]{\includegraphics[width=0.5\textwidth]{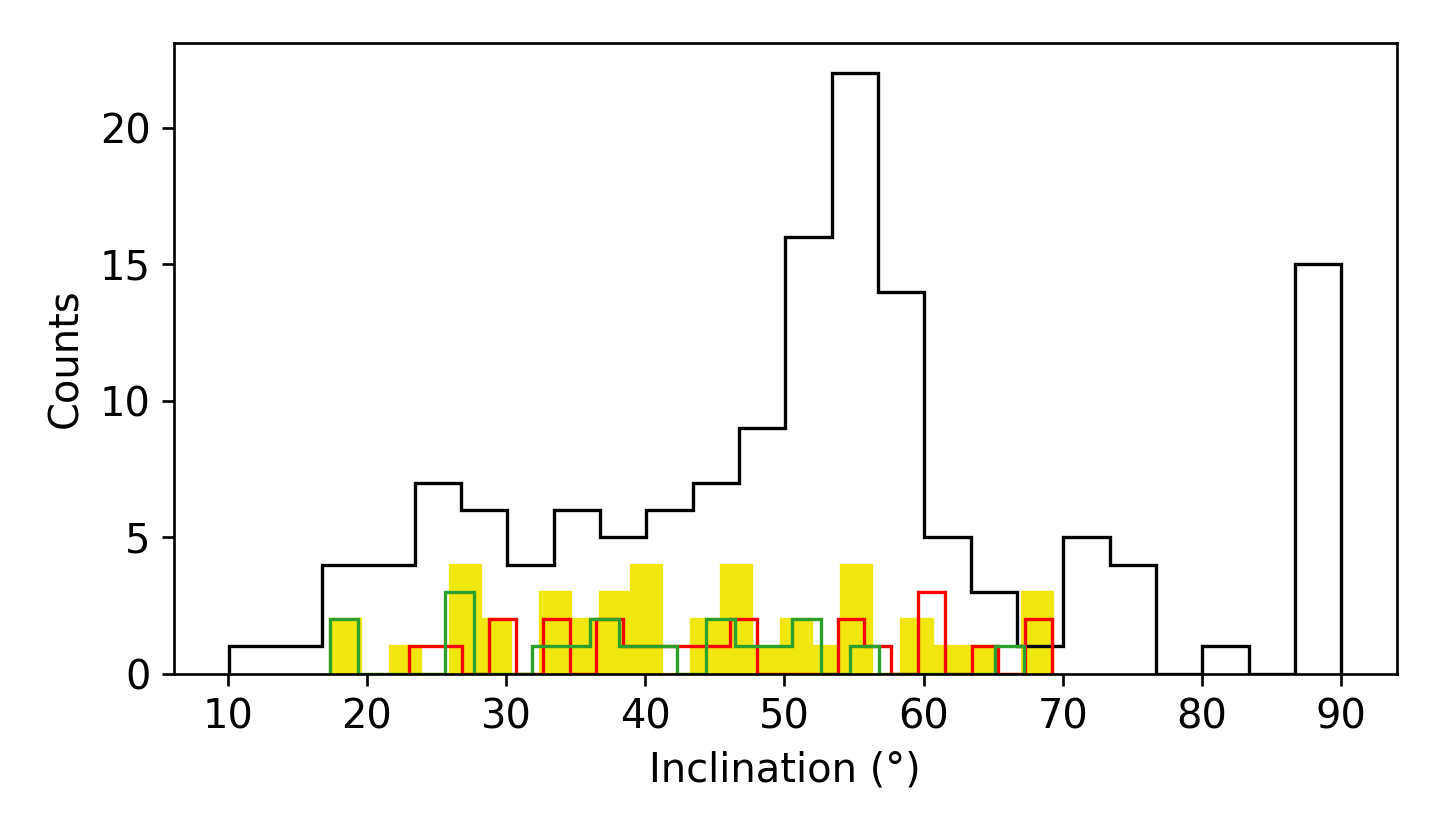}}
	\subfloat[]{\includegraphics[width=0.5\textwidth]{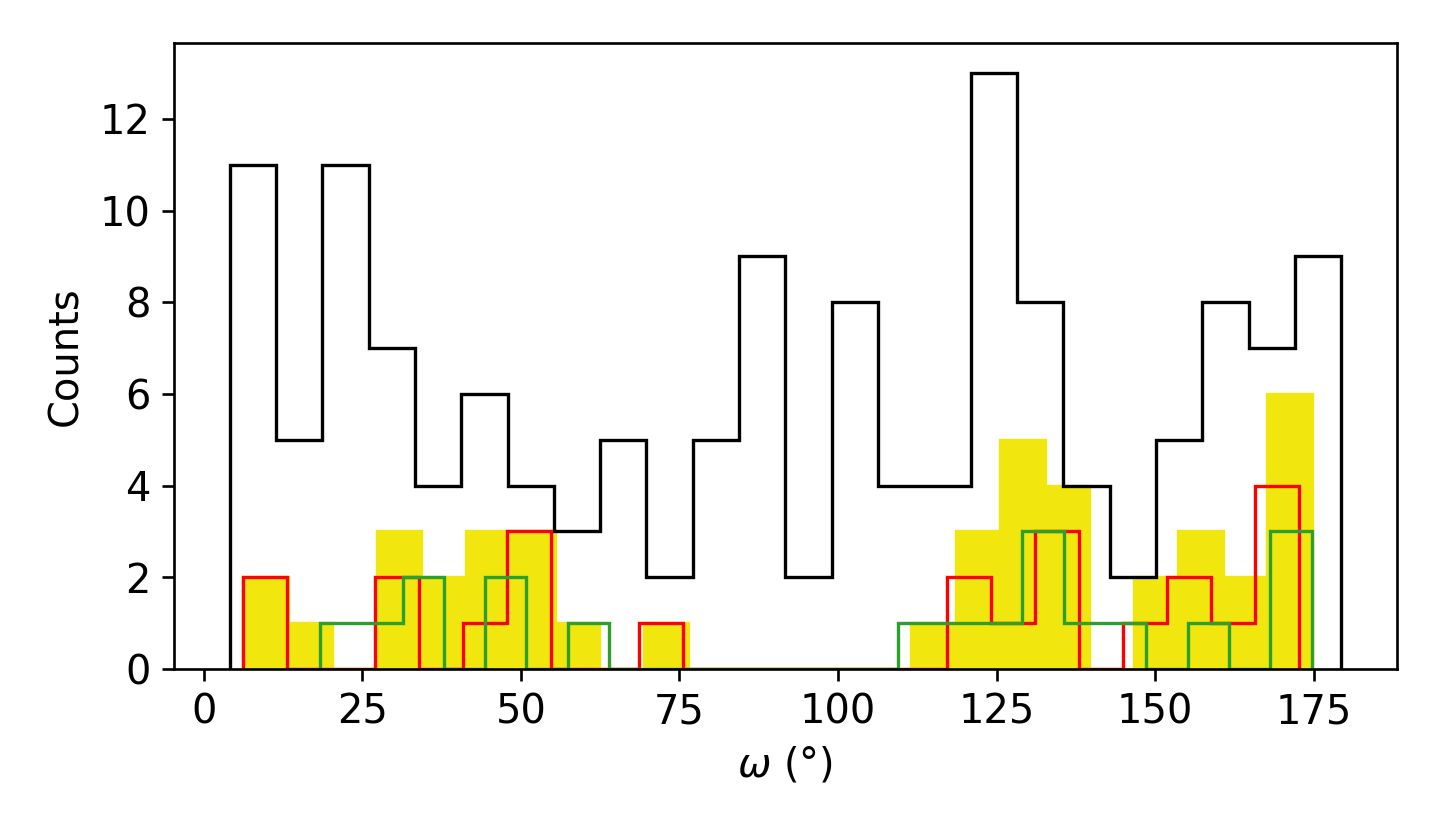}}\\
	\subfloat[]{\includegraphics[width=0.5\textwidth]{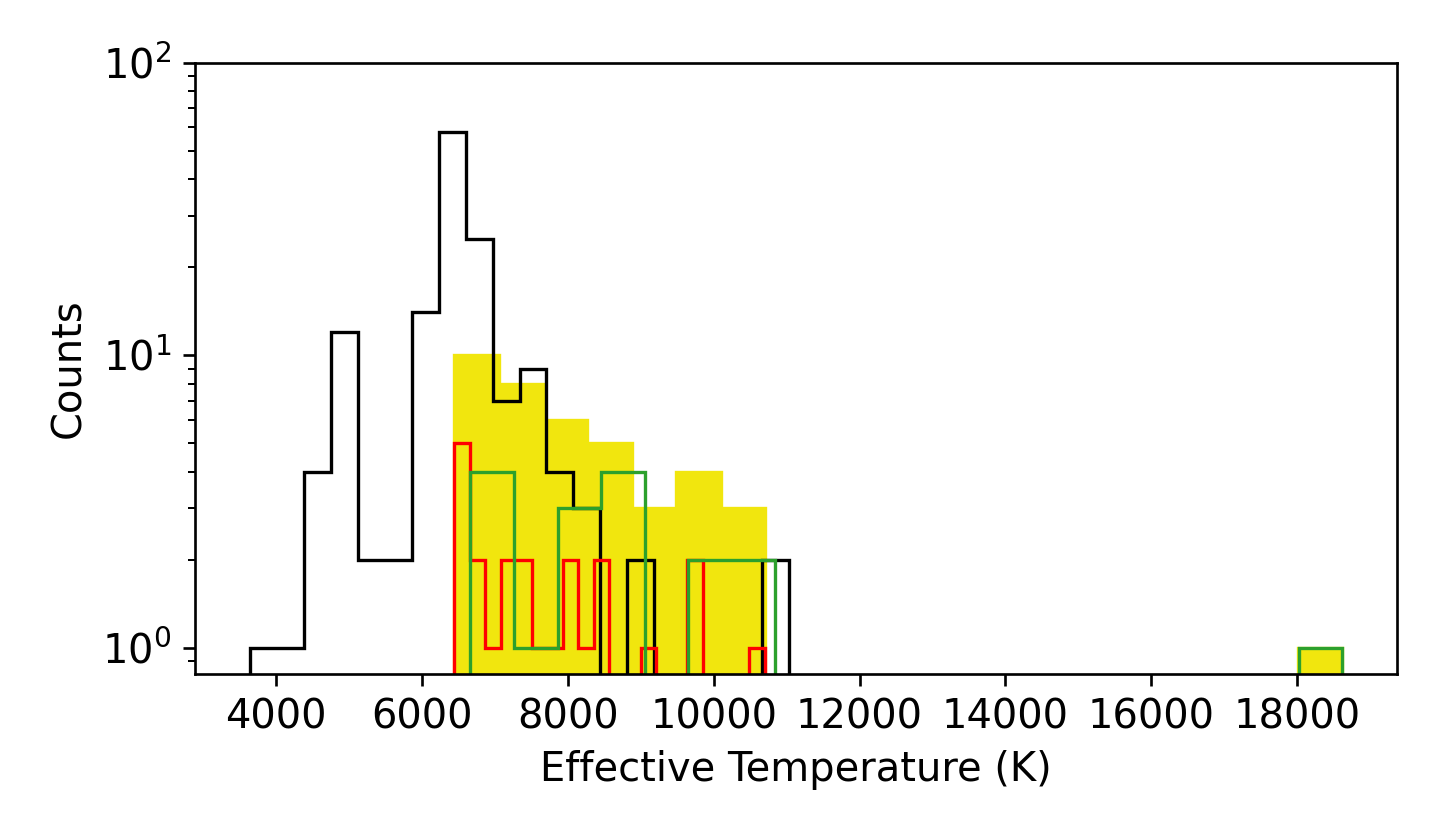}}
	\subfloat[]{\includegraphics[width=0.5\textwidth]{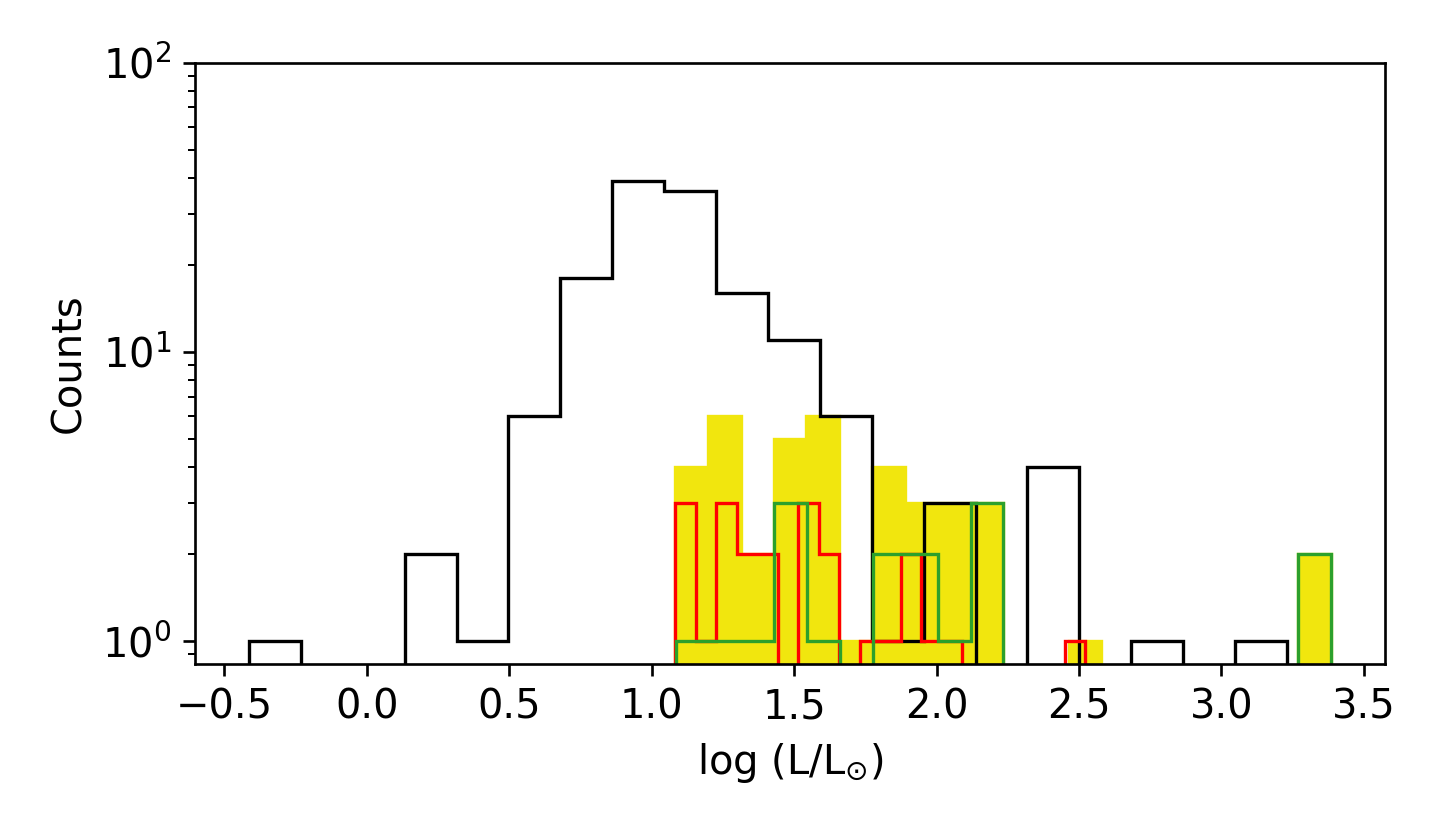}}\\
	\caption{Histograms of the parameters for the HBSs. Black curves represent the HBSs from \citetalias{2023ApJS..266...28L}. Green curves represent the HBSs from \citetalias{2021A&A...647A..12K}. Red curves represent the 23 HBSs in this work. The yellow areas represent all TESS HBSs (this work and \citetalias{2021A&A...647A..12K}). The period, eccentricity, inclination, argument of periastron ($\omega$), effective temperature, and luminosity are in panel (a)$-$(f), respectively.
		\label{fig:dis}}
\end{figure*}

\subsection{Additional Variations in TIC 265473090}
As can be seen in panel (c) of Figure \ref{fig:hbs2}, TIC 265473090 does not seem to fit well. There is a distinct minima outside the heartbeat signal, and the additional variations are also present in the residuals panel. A similar feature also occurs in OGLE-LMC-HB-0287 reported by \citet{2022ApJS..259...16W} (see their Figure 8). We suggest that such variability is caused by a dark spot on the surface of a component. Assuming that the component rotates pseudo-synchronously with the orbital period, the observer will see a dimming caused by the spot and a heartbeat variation near the periastron passage. The observed light curves are then a combination of these two effects. However, it is unclear whether the dark spot is on the primary or secondary surface.

\section{Summary and Conclusions}\label{sec:conclusions}
In this work, we report 23 new non-eclipsing HBSs based on TESS data \footnote{While this paper was under review, we noticed that \citet{2024arXiv240714421S} reported 180 TESS HBSs, four of which overlap with ours, including the TICs 137810570, 265473090, 370209445, and 370269453.}. We also derive the orbital parameters of these HBSs using the K95$^+$ model based on the MCMC method. The suitability of the K95$^+$ model for analyzing non-eclipsing HBSs \citep{2023ApJS..266...28L} lends credence to the reliability of the derived parameters in this study. The e$-$P relation of these objects also shows a positive correlation \citep{2023ApJS..266...28L}, and the existence of orbital circularization \citep{2012ApJ...753...86T, 2016ApJ...829...34S}. The H-R diagram shows that the HBSs can appear in a much wider area. The distribution of the parameters indicates that the TESS survey is more likely to detect HBSs with shorter periods. Furthermore, the more massive and higher temperatures and luminosities of these samples may be due to selection effects.

The lowest eccentricity system among these objects is TIC 118024242 with an eccentricity of 0.08. Another Kepler HBS with a low eccentricity of 0.03 is KIC 10096019 \citep{2023ApJS..266...28L}. Although the eccentricity of most HBSs is greater than 0.2, the ``heartbeat'' feature of the light curves suggests that HBSs may have eccentricities less than 0.1.

Tidally excited oscillation (TEO) \citep{2016ApJ...829...34S, 2017MNRAS.472.1538F} is an important feature of HBS. However, no TEO is detected in these samples, and other HBSs with TEO in this series are in preparation.

There are some HBSs that are reported as triple systems \citep{2016MNRAS.463.1199H, 2021MNRAS.508.3967O}. The fractional flux offset $C$ in Equation (\ref{equation:one}) already includes contributions from the third light in some sense. However, it is challenging to distinguish the fraction of the third light from other physical effects. On the other hand, \citet{2016MNRAS.463.1199H} have provided a practical way to detect whether a third body is present in non-eclipsing HBSs. They have found that the observed apsidal motion rate has a large deviation from theoretical expectations, and have concluded that a third body is present in their sample. This will be an appropriate approach for future studies of the third body in these HBSs.

Although we have discovered a small percentage of HBSs in this work, we believe that there are still a large number of HBSs to be found in the TESS data, as the relatively few HBSs based on TESS data are currently reported. We aim to discover more TESS HBSs in our future work.

\section*{Acknowledgements}

This work is partly supported by the International Cooperation Projects of the National Key R$\&$D Program of China (Grant No. 2022YFE0127300), the National Natural Science Foundation of China (Nos. 11933008 and 12103084), the Basic Research Project of Yunnan Province (Grant Nos. 202201AT070092 and 202301AT070352), the Science Foundation of Yunnan Province (No. 202401AS070046), and the Yunnan Revitalization Talent Support Program. Funding for the TESS mission is provided by the NASA Explorer Program. The Gaia Survey is a cornerstone mission of the European Space Agency. We thank the TESS and Gaia teams for their support and hard work. We are grateful to the anonymous referee for many heuristic comments that helped to improve this manuscript.

%%%%%%%%%%%%%%%%%%%%%%%%%%%%%%%%%%%%%%%%%%%%%%%%%%
\section*{Data Availability}
The data underlying this article are available at the Mikulski Archive for Space Telescopes (MAST) (\url{https://mast.stsci.edu/}), or can be downloaded using the {\tt\string lightkurve} package \citep{2018ascl.soft12013L}.}
 
%%The inclusion of a Data Availability Statement is a requirement for articles published in MNRAS. Data Availability Statements provide a standardised format for readers to understand the availability of data underlying the research results described in the article. The statement may refer to original data generated in the course of the study or to third-party data analysed in the article. The statement should describe and provide means of access, where possible, by linking to the data or providing the required accession numbers for the relevant databases or DOIs.


\bibliographystyle{mnras}
\bibliography{example} % if your bibtex file is called example.bib

%\begin{thebibliography}{99}
%\end{thebibliography}

%%%%%%%%%%%%%%%%%%%%%%%%%%%%%%%%%%%%%%%%%%%%%%%%%%

%%%%%%%%%%%%%%%%% APPENDICES %%%%%%%%%%%%%%%%%%%%%
%%\onecolumn
%%\appendix

%%%%%%%%%%%%%%%%%%%%%%%%%%%%%%%%%%%%%%%%%%%%%%%%%%

% Don't change these lines
\bsp	% typesetting comment
\label{lastpage}
\end{document}